\documentclass[12pt,a4paper]{article}
\pdfoutput=1
\usepackage{hyperref}
\usepackage{amsmath}
\usepackage{amssymb}
\usepackage{mathrsfs}
\usepackage{bbm}
\usepackage{graphicx}
\usepackage{color}
\usepackage{enumerate}
\usepackage[normalem]{ulem}

\numberwithin{equation}{section}

\definecolor{darkgreen}{rgb}{0,.5,0} \definecolor{darkred}{rgb}{.7,0,0}

\begin{document}

\title{Phases of Holographic Hawking Radiation on spatially compact spacetimes}

\author{Donald Marolf${}^{\sharp}$ and Jorge E. Santos${}^{\diamond}$\\
  \\
  {\it ${}^\sharp$Department of Physics, University of California at Santa Barbara,}\\
    {\it Santa Barbara, CA 93106, U.S.A.} \\
  \\
  {\it ${}^\diamond$Department of Applied Mathematics and Theoretical Physics,}\\
    {\it University of Cambridge, UK}
  }

\date{\today}

\maketitle

\begin{abstract}
We study phases of equilibrium Hawking radiation in $d$-dimensional holographic CFTs on spatially compact spacetimes with two black holes.  In the particular phases chosen the dual $(d+1)$-dimensional bulk solutions describe a variety of black funnels and droplets.  In the former the CFT readily conducts heat between the two black holes, but it in the latter such conduction is highly suppressed.  While the generic case can be understood in certain extreme limits of parameters on general grounds, we focus on CFTs on specific geometries conformally equivalent to a pair of $d \ge 4$ AdS${}_d$-Schwarzschild black holes of radius $R$.  Such cases allow perturbative analyses of non-uniform funnels associated with Gregory-Laflamme zero-modes.  For $d=4$ we construct a phase diagram for pure funnels and droplets by constructing the desired bulk solutions numerically.  The fat non-uniform funnel is a particular interesting phase that dominates at small $R$ (due to having lowest free energy) despite being sub-dominant in the perturbative regime. The uniform funnel dominates at large $R$, and droplets and thin funnels dominate at certain intermediate values.  The thin funnel phase provides a mystery as it dominates over our other phases all that way to a critical $R_{\mathrm{turn}}$ beyond which it fails to exist.  The free energy of the system thus appears to be discontinuous at $R_{\mathrm{turn}}$, but such discontinuities are forbidden by the 2nd law.  A new more-dominant phase is thus required near $R_{\mathrm{turn}}$ but the nature of this phase remains unclear.
\end{abstract}

\maketitle

\newpage

\tableofcontents



\section{Introduction}
\label{intro}

Understanding the thermodynamics of strongly interacting field theories remains a challenging task.  This is no less the case when the theory is coupled to heat baths provided by black holes, in which context the results describe phases of the associated Hawking radiation.  But for appropriate large $N$ conformal field theories (CFTs), gauge/gravity duality \cite{Maldacena:1997re} provides what one hopes may be a tractable description in terms of semi-classical gravity in asymptotically (locally) anti-de Sitter (AlAdS) spacetimes.  Below we consider the bulk classical limit in cases where the bulk description may be truncated to Einstein-Hilbert gravity with a $\Lambda < 0$ cosmological constant.

Early explorations \cite{AM,Wiseman:2001xt,Wiseman:2002zc,Casadio:2002uv,Karasik:2003tx,Kudoh:2003xz,
Kudoh:2003vg,Kudoh:2004kf,Karasik:2004wk,Yoshino:2008rx} of both this setting and the related context of brane-world black holes found bulk solutions for which the dual CFT Hawking radiation behaved quite differently from that of a free theory.  In particular, despite a large density of CFT states $s_{\mathrm{CFT}}$, the flux of energy to infinity remained quite small (order 1).  However, inspired in part by  \cite{Fitzpatrick:2006cd}, it was later recognized in \cite{Hubeny:2009ru} that such solutions represented only one possible phase of the Hawking radiation with other phases allowing heat transport of the same order as the CFT entropy density $s_{\mathrm{CFT}}$. Thus the CFT undergoes a conducting/insulating phase transition that is related by a conformal transformation to the more familiar confinement/deconfinement transition \cite{Marolf:2013ioa}.

The properties of the above phases are readily seen from the dual $d+1$-dimensional AlAdS bulk solutions. The induced conformal metrics on their conformal boundaries must all agree with that of the $d$-dimensional black hole spacetime on which the CFT is defined.  Following \cite{Hubeny:2009ru}, we thus refer to the CFT heat baths as boundary black holes.  There will also be one or more black holes in the bulk whose horizons end on those of the boundary black holes. Bulk horizons conduct heat along themselves at the classical level \cite{Thorne:1986iy} and thus at a level proportional to the CFT entropy of states $s_{\mathrm{CFT}}$.  But two disconnected bulk horizons exchange heat only via bulk Hawking radiation, which remains an order 1 effect at large $s_{\mathrm{CFT}}$.  The basic thermal conductivity properties of the CFT are thus determined by the pattern of bulk connections between the various boundary black holes.  This argument, presented in \cite{Hubeny:2009ru}, has been verified by direct studies of connected \cite{Fischetti:2012ps,Figueras:2012rb,Fischetti:2012vt} and isolated \cite{Haehl:2012tw} boundary black holes. See \cite{Emparan:2013fha} for a simple solvable example describing heat transport along an analogous horizon.

Bulk horizons connecting two or more boundary black holes have become known as black funnels, while bulk horizons that connect to only one boundary black hole are called black droplets \cite{Hubeny:2009ru}.  In this terminology it is standard to treat any asymptotic region of a globally hyperbolic CFT spacetime as an additional black hole, and thus to also apply the term funnel to bulk horizons that connect boundary black holes to the asymptotic regions of the CFT.  Such asymptotic regions are in fact conformally equivalent to black holes in simple cases (see \cite{Fischetti:2013hja}, based on \cite{Headrick:2010zt,Marolf:2010tg,Hung:2011nu}).  More generally, the asymptotic regions of globally hyperbolic CFT spacetimes can be mapped to spacetimes with null singularities which one might consider to be singular black holes.

\begin{figure}
\centerline{
\includegraphics[width=0.4\textwidth]{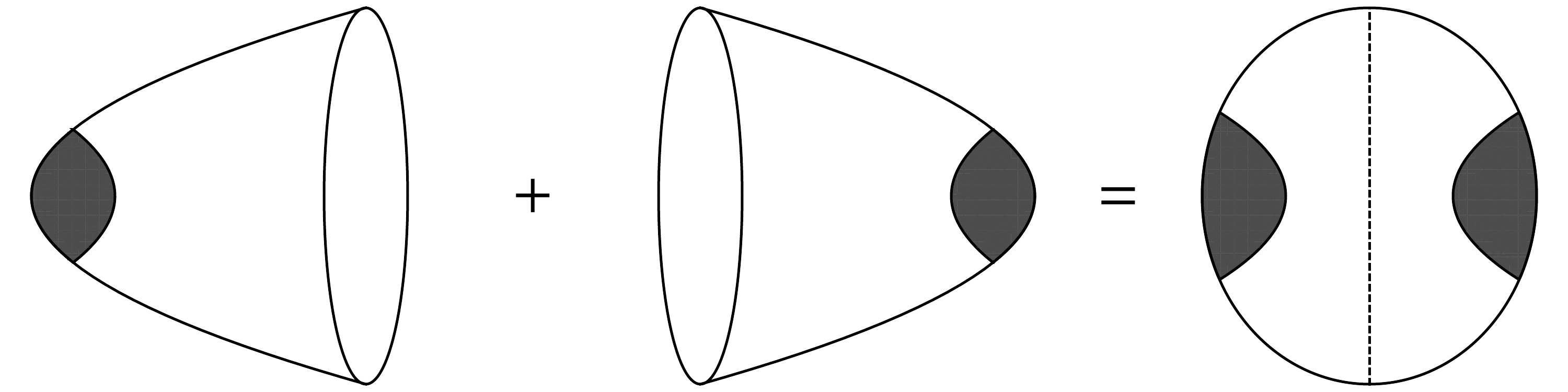}
}
\caption{Two copies (left) of conformally-compactified Schwarzschild-AdS${}_d$ may be glued together to make a spacetime (right) describing a pair of black holes (right) on the Einstein Static Universe. Static slices are shown. }
\label{fig:bndy}
\end{figure}

Our purpose below is to study phase transitions between AlAdS${}_{(d+1)}$ funnels and droplets associated with a particular class of boundary black hole spacetimes conformally equivalent to a pair of (global) $d \ge 4$ Schwarzschild-AdS${}_d$ glued together along the AdS boundary. See figure \ref{fig:bndy} and section \ref{pert} for technical details.  The resulting boundary spacetime is conformally equivalent to a pair of black holes in the Einstein Static Universe (ESU), generalizing the familiar way in which the ESU $S^{d-1} \times {\mathbb R}$ is conformally equivalent to two copies of AdS${}_d$ and gives the conformal boundary of global AdS${}_{d+1}$.   Thus we consider what one might call global droplets and global funnels.  The analogous phase transitions were studied analytically for $d=3$ with Ba\~nados-Teitelboim-Zanelli (BTZ) boundaries in \cite{Hubeny:2009rc}, where it was argued that all relevant phases for this case can be constructed by double Wick rotation of global AdS${}_4$ and Schwarzschild-AdS${}_4$.

While a great many $d \ge 4$ funnel and droplet solutions have by now been constructed \cite{Gregory:2008br,Hubeny:2009kz,Hubeny:2009rc,Caldarelli:2011wa,
Figueras:2011va,Figueras:2011gd,Santos:2012he,Haehl:2012tw,Fischetti:2013hja,Figueras:2013jja,Santos:2014yja,Emparan:2015hwa,Mefford:2016res,Fischetti:2016oyo},
our work is the first to construct multiple such phases for the same boundary black holes.  This allows a meaningful comparison of their free energies.  In particular, although both droplet and funnel free energies $F = E-TS$ receive divergent contributions from the infinite area of the non-compact bulk horizons,
the fact that both solutions satisfy the same boundary conditions and that the AlAdS boundary is spatially compact implies that the difference $\Delta F$ must be finite and unambiguous as defined using any Fefferman-Graham regulator\footnote{Note that $\Delta F$ can be computed from the Euclidean action, which is manifestly finite after holographic renormalization.  Furthermore, for any boundary dimension $d$ $\Delta F$ will be independent of the choice of boundary conformal frame since any conformal anomaly that might afflicts the definition of $F$ for either solution will cancel in computing $\Delta F$.}.   We consider equilibrium situations in which all bulk horizons are at the same temperature $T$, which we take to agree with the temperatures of all boundary black holes.   It should be noted that more general `detuned' equilibrium solutions should also exist, where the temperatures of the bulk and boundary black holes differ. Such solutions were predicted in \cite{Fischetti:2012vt} based on \cite{Headrick:2010zt,Marolf:2010tg,Hung:2011nu} and constructed with specific boundary metrics in \cite{Fischetti:2016oyo}. As we discuss in section \ref{pert}, the appearance of the Gregory-Laflamme instability in 5 or more bulk dimensions (\emph{i.e.}, for $d \ge 4$) leads to additional funnel phases not seen in the $d=3$ analysis of \cite{Hubeny:2009rc}.

We begin with a brief overview in section \ref{overview} of the phases to be expected for general ESU boundary black holes. Here we simply characterize the boundary spacetime by the size $R$ and temperature $T$ of the boundary black hole and the ESU length scale $\ell_d$, describing the phases to be expected in various extreme limits of the dimensionless parameters $R/\ell_d$ and $T\ell_d$.  This generalizes the general discussion of phases in \cite{Hubeny:2009ru}.  Further insight into droplet phases is then obtained  by recalling the $d=3$ analysis of \cite{Hubeny:2009rc} (section \ref{BTZ}), and insight into funnel phases is obtained by considering the Gregory-Laflamme instability (section \ref{pert}) with some detailed expressions relegated to appendix \ref{ap:1}.   Numerical methods and the framework for our calculations are described in section \ref{numerics} while diagnostic machinery is described in section \ref{sec:diag}.  Results for $d=4$ are  presented in section \ref{sec:results} and are supported by convergence tests described in appendix \ref{ap:2}.  We close with further discussion in section \ref{sec:disc}. In particular, an argument based on the second law of thermodynamics predicts that a further new phase must exist with lower free energy must exist in certain regions of parameter space, but we find no natural candidates for this phase.

\section{Dominant Funnels and Droplets for general ESU black hole boundaries}
\label{overview}

We begin by considering general compact static boundary spacetimes containing a pair of boundary black holes.    We suppose here that the black holes lie on opposite sides of the boundary spacetime as shown in figure \ref{fig:bndy}, though it is also interesting to consider phase transitions that occur as one varies the relative separation of the two black holes.  We first discuss the relevant notion of the thermodynamic 2nd law in section \ref{2nd} to establish that the most thermodynamically stable solutions are those with lowest $\Delta F$.  We then combine general arguments from \cite{Hubeny:2009ru} with observations from studies of the Gregory-Laflamme instability in section \ref{DFgen} to motivate the rough form of a general phase diagram for droplets and funnels.  Finally, we review analytic results \cite{Hubeny:2009rc} for $d=3$ with BTZ boundaries in section \ref{BTZ}.

\subsection{The second law for droplets and funnels}
\label{2nd}

Before discussing the various possible phases, note that for general boundary metrics one may expect a variety of phases to exist, for which one will thermodynamically dominate over the others.
Recall that we study a system in equilibrium with a heat bath provided by the black hole on the boundary.  One thus expects the relevant notion of dominance to be determined by minimizing the free energy $F = E-TS$ defined by the heat bath temperature $T$; \emph{i.e.}, where $T$ is fixed by the surface gravity of the boundary black hole.  Indeed, for Lorentzian AlAdS spacetimes with such boundaries it was shown in \cite{Bunting:2015sfa} that finite processes cannot decrease the renormalized free energy $F$ defined by using the entropy of the bulk event horizon. The same must hold for our $\Delta F$ since it is defined by subtracting off the free energy of a fixed reference solution.  This gives a useful form of the 2nd law for our systems.

This point is sufficiently important that it is useful to give an expanded discussion over the next few paragraphs.  Many readers will find it most clear to consider the relevant thermodynamics from the viewpoint of a dual CFT living on the boundary spacetime.  This boundary spacetime contains black holes.  CFT entropy can disappear into the black hole, can also be emitted from the black holes, so the CFT outside the black hole is not a closed system and no law of thermodynamics can forbid $\Delta S$ from decreasing.

In contrast, if the black holes were dynamical, then we should find the {\it total} entropy $S_{\mathrm{total}} = S_{\mathrm{CFT}} +  S_{\mathrm{bndy}\ \mathrm{BHs}}$ to be non-decreasing.  That is to say, if we made gravity dynamical on the boundary, then there would be a boundary Generalized Second Law (GSL).

However, the boundary metric is not dynamical.  We can think of this as taking the limit of a theory with dynamical boundary metric (with some boundary Newton constant $G_{\mathrm{bndy}}$) and in particular taking $G_{\mathrm{bndy}} \rightarrow 0$.  In this limit, the Bekenstein-Hawking entropy of the boundary black holes diverges and, moreover, the boundary geometry does not change as energy flows in and out of the boundary black holes.  The first fact makes the boundary GSL useless in the standard form (non-decrease of $S_{\mathrm{CFT}} +  S_{\mathrm{bndy}\ \mathrm{BHs}}$),  but the second fact comes to the rescue.  Since changes in the boundary geometry are tiny, we can use the 1st law for boundary black holes to write $\mathrm{d}S_{\mathrm{bndy}\ \mathrm{BHs}} = \mathrm{d}E_{\mathrm{bndy}\ \mathrm{BHs}}/T_{\mathrm{BHs}}$.  But energy conservation would also guarantee that $\mathrm{d}E_{\mathrm{bndy}\ \mathrm{BHs}} = - \mathrm{d}E_{\mathrm{CFT}}$, so the boundary GSL in fact forbids decreases in the quantity
\begin{equation}
S_{\mathrm{CFT}} - E_{\mathrm{CFT}}/T_{\mathrm{BH}} = -F_{\mathrm{CFT}}/T_{\mathrm{BH}}.
\end{equation}
That is to say, it states that the free energy $F_{\mathrm{CFT}}$ must be non-increasing.  This is just the usual way in which the 1st law allows us to rewrite the closed-system 2nd law as a useful 2nd law for open systems interacting with a heat bath.  The above argument was the primary motivation for \cite{Bunting:2015sfa}, which showed that the corresponding bulk spacetimes do indeed satisfy such an (open system) version of the 2nd law.

\begin{figure}
\centerline{
\includegraphics[width=0.4\textwidth]{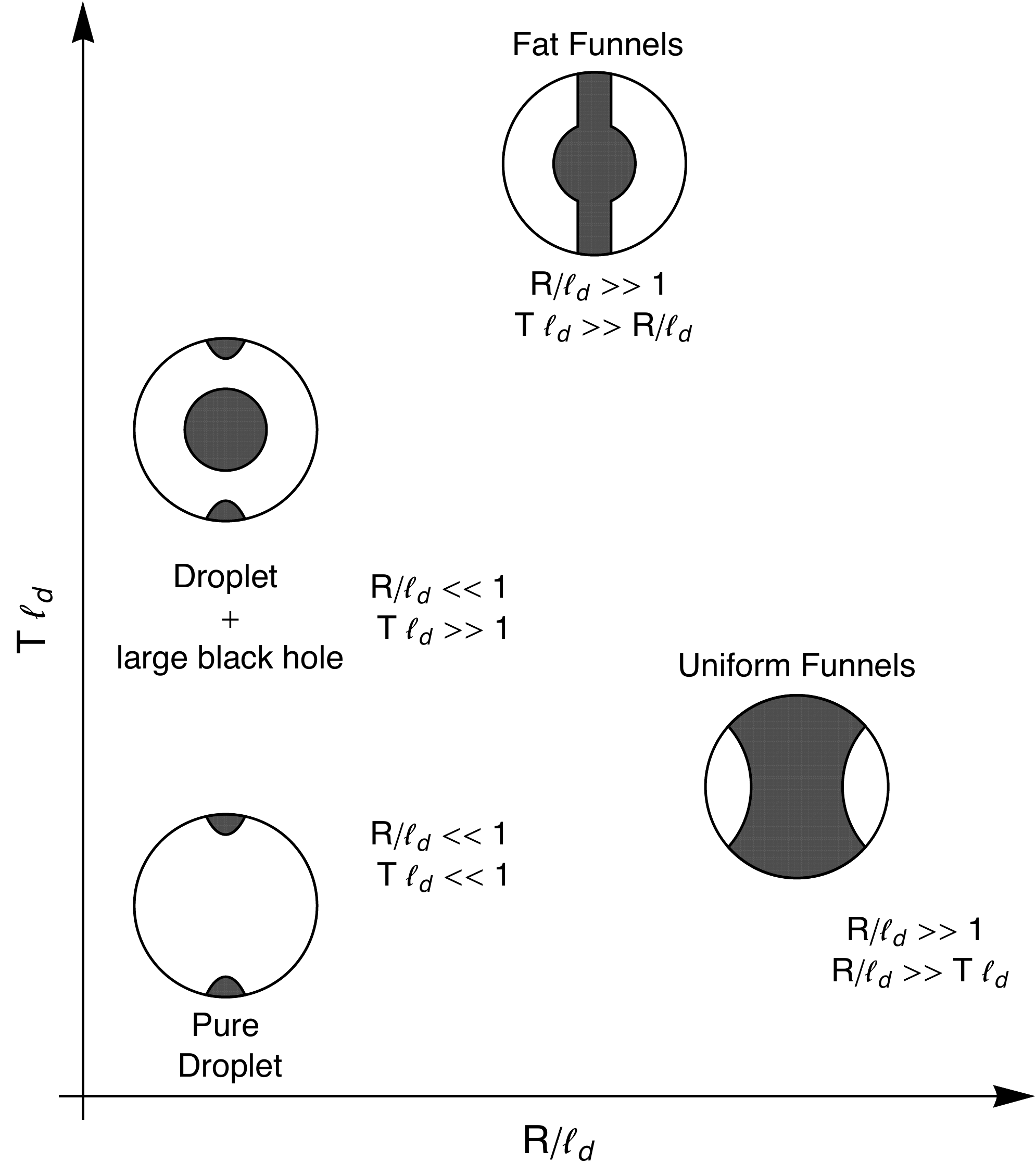}
}
\caption{Phase diagram for general $R/\ell_d, T \ell_d$. The boundaries between phases are not shown as their locations are unknown.  We cannot rule out the possibility that additional phases (such as thin funnels) also dominate in certain regimes, though comparison with known results for Kaluza-Klein black holes \cite{Kol:2002xz,Sorkin:2003ka,Kudoh:2003ki,Kudoh:2004hs,Kol:2005vy,Headrick:2009pv,Kalisch:2015via,Kalisch:2016fkm,Kalisch:2017bin}.  makes this seem unlikely.  The absence of a Gregory-Laflamme instability for $d=3$ boundary dimensions suggests that in that case the fat and uniform funnels are connected by a crossover, though in higher dimensions we expect a sharp phase transition.  }
\label{fig:Full}
\end{figure}
\subsection{Droplets and Funnels for general boundary black holes}
\label{DFgen}

We are now ready to discuss natural phases of droplets and funnels for general boundary black holes. As argued in \cite{Hubeny:2009ru}, one expects the radius $R$ and the temperature $T$ of the boundary black holes to be important in determining both allowed and dominant phases.  The spatial scale $\ell_d$ of the compact boundary will also play a role.  Note that when the boundary spacetime is conformally equivalent to a pair of AdS${}_d$ black holes, this $\ell_d$ is also the boundary-AdS scale. The details of the associated transformation will be reviewed in section \ref{pert}.

As we explain below, the arguments of \cite{Hubeny:2009ru} generalize readily to our current context and suggest the rough phase diagram shown in figure \ref{fig:Full}. For the moment we consider general independent $R, \ell_d, T$, though for Schwarzschild-AdS or BTZ boundary black holes the dimensionless combinations are all determined by a single parameter (e.g., $R/\ell_d$).  Let us begin to understand figure \ref{fig:Full} by analyzing the limit $R/\ell_d \rightarrow 0$ at fixed $T\ell_d$ (\emph{i.e.}, along the far left side).  Since the boundary black holes are then tiny compared to all other scales, they naturally support bulk horizons that extend only slightly into the bulk.  This suggests a droplet phase similar to that of \cite{Figueras:2011gd} near each droplet for small $R/\ell_d$.  As we will recall in section \ref{BTZ}, the $d=3$ analysis of \cite{Hubeny:2009rc} suggests that there can be more than one droplet solution for given boundary conditions but that droplets whose horizons stay closest to the boundary will dominate.  Such solutions are called `short' droplets in section \ref{BTZ} and we discuss only such droplets here.

The droplets in this regime make only minimal impact on the bulk physics.  Thus the phase structure along the left side of figure \ref{fig:Full} must be otherwise identical to that in the absence of boundary black holes and is dictated by the Hawking-Page transition \cite{Hawking:1982dh}.  At large $T \ell_d$, the dominant phase will contain two droplets as well as a large central AdS-Schwarzschild-like black hole.  This phase is dynamically stable because the AdS gravitational potential inhibits attempts to merge the central black hole with either droplet. There will also be dynamically unstable phases involving more bulk black holes, carefully balanced between the central black hole and the droplets.  But of greater interest are two dynamically stable sub-dominant phases:  a pure droplet phase with no central black hole, and a phase having both droplets and a small central black hole.  In the limit $R/\ell_d \rightarrow 0$, the transition occurs at the Hawking-Page temperature $T_{\mathrm{HP}} = (d-2)/(2\pi \ell_d)$. At lower temperatures the pure droplet phase dominates, and at sufficiently low temperatures it is the only phase that exists (see figure \ref{fig:Full}).

We will now probe larger values of $R/\ell_d$ at high temperature, continuing clockwise around figure \ref{fig:Full}. As measured by a standard dimensionless Fefferman-Graham coordinate $z$ (with $z=0$ on the boundary), we expect the droplet to penetrate a distance $z \sim R/\ell_d$ at small $R/\ell_d$.  This is a consequence of scale/radius duality (and thus of the bulk symmetries in empty global AdS).  But the $z$-location of the large central black hole's horizon is set by the temperature, with $z \sim T \ell_d$.  So for $RT \gg 1 $ any large central black hole should merge with the droplets to form a funnel. The change in horizon topology requires some sort of phase transition, though the details remain to be investigated.  If we are still at $R/\ell_d  \ll T \ell_d$ then, as measured by natural coordinates on the compactified spacetime, the droplets remain thin relative to the black hole and the resulting funnel will display the large bulge shown in left-most part of this region of the phase diagram. The solution will be well-approximated by the global bulk Schwarzschild-AdS black hole (with ESU boundary) over most of the spacetime, with significant departures only very close to the poles of the ESU.  We call this the fat funnel phase, though at small $R/\ell_d$ this need not imply the existence of other funnels; it is possible that all sub-dominant phases are droplets. On the other hand, thinking of a fat funnel as in some sense composed of a more uniform funnel and a large black hole, at large $R/\ell_d$ (with still $T \ell_d \gg R/\ell_d$) the Hawking-Page transition for pure ESU boundaries suggests the presence of (at least) two further funnel phases (subdominant at large $T\ell_d$), which we call thin and uniform.

Increasing $R/\ell_d$ at fixed $T \ell_d$ will gradually change the shape of the funnel to make the central bulge less pronounced.  For large $R/\ell_d$ and small $T \ell_d$, one expects the funnel to reach down a distance $\Delta z \sim \ell_d/R$ into the bulk.  So the central bulge should disappear completely in the limit $R/\ell_d \gg T \ell_d$.   As we will review in section \ref{pert} below, an analogy with the Gregory-Laflamme instability and the analysis of \cite{Gubser:2001ac,Sorkin:2004qq} suggests that for $d \ge 4$ there is in fact a sharp phase transition from fat funnels to more uniform funnels, though the lack of a Gregory-Laflamme instability for $d=3$ predicts a cross-over in that case.   We will identify this phase transition below for $d=4$ AdS-Schwarzschild boundary black holes.  The uniform funnel phase should then persist as we decrease $T \ell_d$ keeping $R/\ell_d$ sufficiently large.  At small $T \ell_d$, decreasing $R/\ell_d$ will result in a transition back to the pure droplet phase.

This completes our discussion of the expected phase structure for general $R, \ell_d, T.$  Although the dominant phases are clear in various asymptotic regimes, the details of the transitions and many features of possible sub-dominant phases remain to be explored.  For example, both droplets and funnels may persist well into regimes where the other phase dominates.

For the rest of this work we therefore restrict attention to a particular 1-parameter family of boundary black hole metrics for each $d$. Section \ref{BTZ} will briefly review the $d=3$ results of \cite{Hubeny:2009rc} using BTZ boundaries in order to provide further insight into the droplet phases.  This structure seems likely to persist for general black holes with $d \ge 4$. Section \ref{pert} then considers $d \ge 4$, taking the $d$-dimensional boundary metric to be given by a pair of Schwarzschild-AdS${}_d$ (SAdS) black holes.  We focus in particular on insights from the Gregory-Laflamme instability, which introduce qualitative differences from $d=3$.  Since in both cases the boundary metrics are labeled by a single free parameter, not all of the above phases need arise.

\begin{figure}
\centerline{
\includegraphics[width=0.4\textwidth]{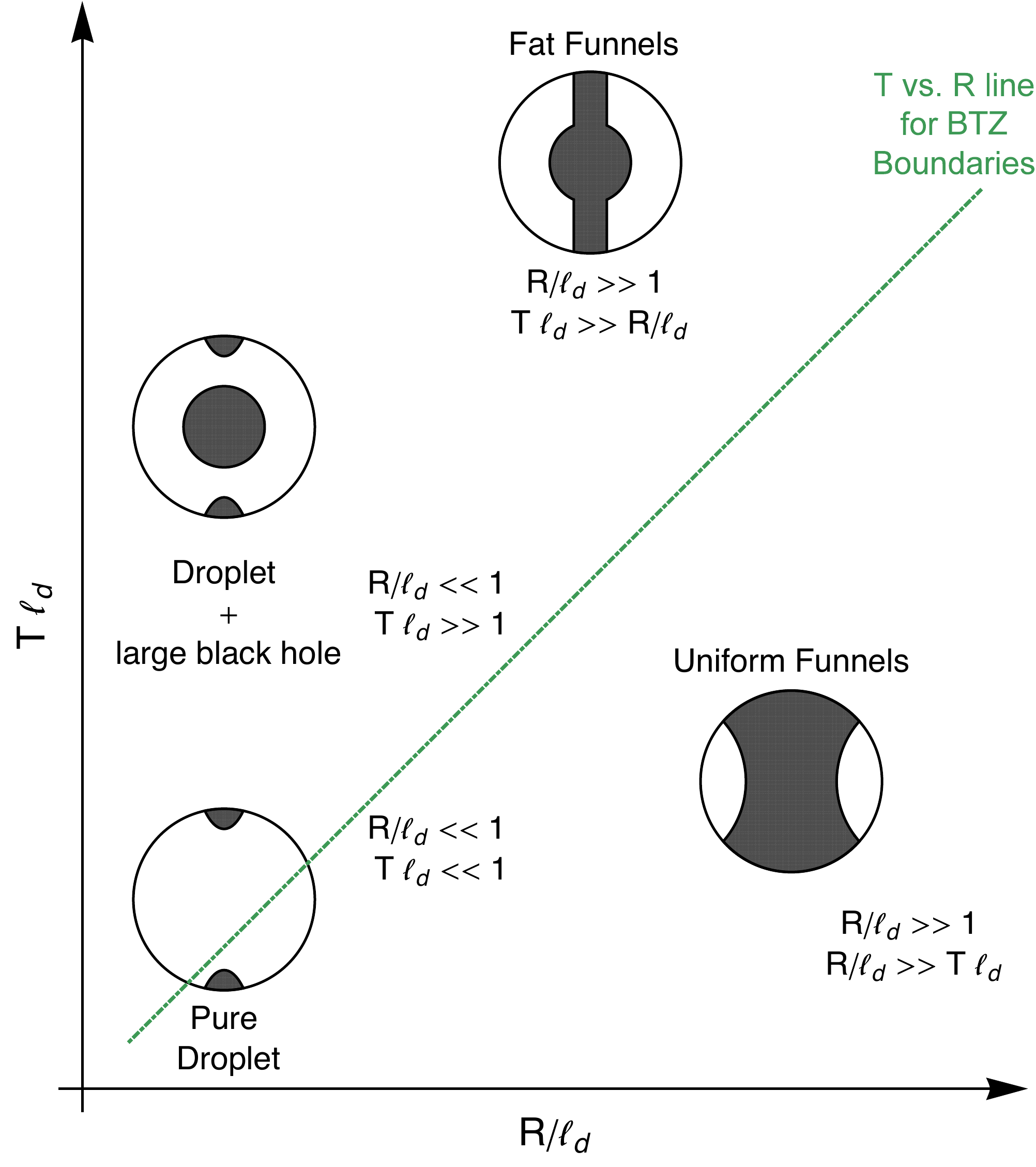}\hspace{1.5 cm}\includegraphics[width=0.4\textwidth]{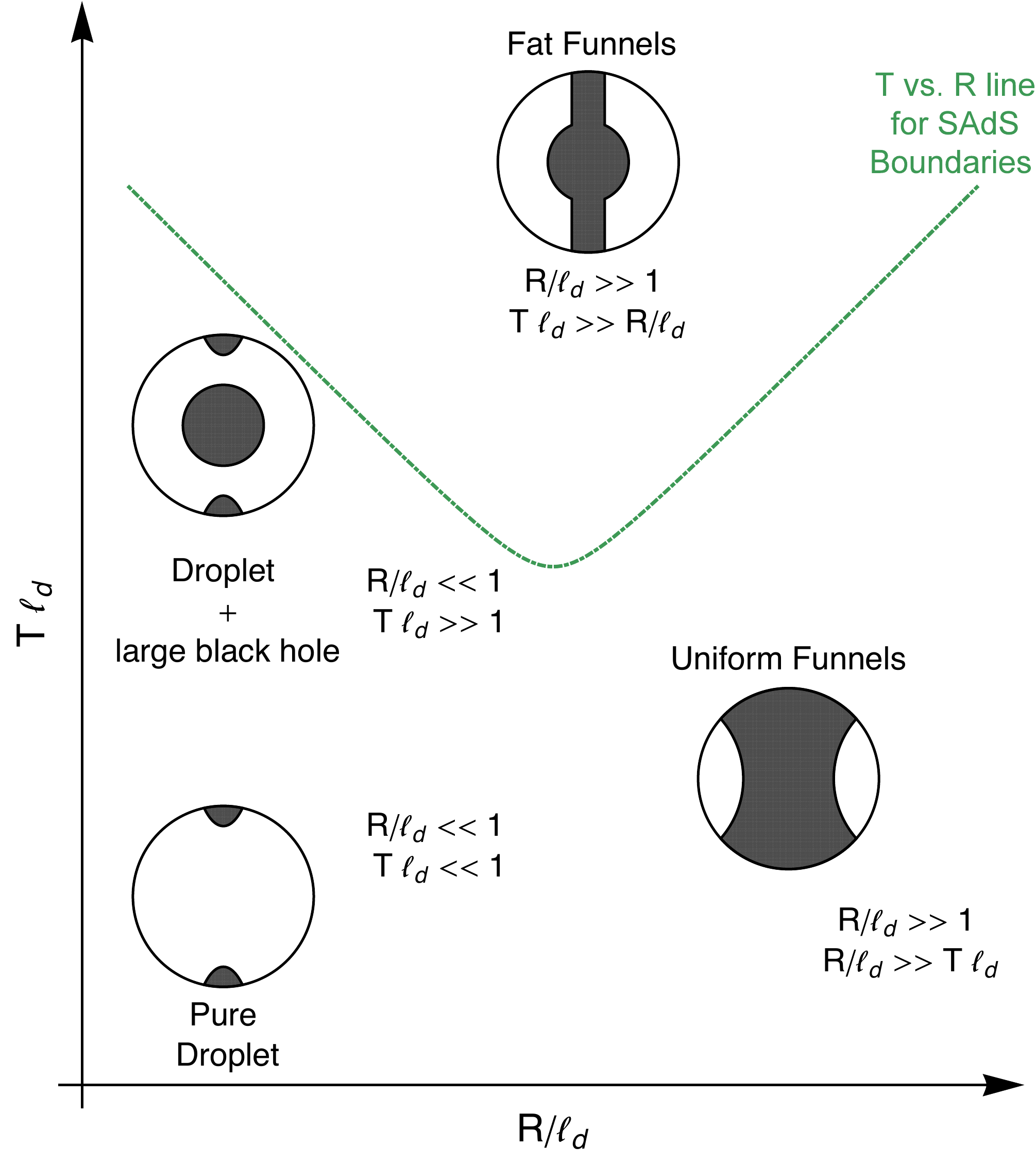}
}
\caption{Dashed lines showing the one-parameter families of boundary conditions studied in \cite{Hubeny:2009rc} (reviewed in section \ref{BTZ}) for $d=3$ (left) and below for $d=4$ (right) as plotted on the phase diagram from figure \ref{fig:Full}.  The boundary metrics studied in \cite{Hubeny:2009rc} contain BTZ black holes while those studied below are Schwarzschild Anti-de Sitter (SAdS).  Recalling that fat and uniform funnels should be connected by a cross-over in $d=3$, the left diagram predicts a transition from droplets at small $R,T$ to funnels at large $R,T$. Since the droplet-plus-black-hole phase is beyond the scope of this work, the right diagram predicts that we will find fat funnels to dominate at small $R$ and that either fat or uniform funnels will dominate at large $R$.  Other phases may of course be possible at intermediate values of parameters.
}
\label{fig:BHlines}
\end{figure}

The relevant curves through our phase diagram are drawn in figure \ref{fig:BHlines}.  The curves lie largely in regions with intermediate values of parameters where the dominant phase is not yet clear.
However, recalling that fat and uniform funnels should be connected by a cross-over in $d=3$, the left diagram for BTZ boundaries predicts a transition from droplets at small $R,T$ to funnels at large $R,T$.  While additional intermediate phase transitions are possible in principle, they would require a special symmetry that arises in this case to be spontaneously broken (see \cite{Hubeny:2009rc} or the review in section \ref{BTZ} below).
And since the droplet-plus-black-hole phase is beyond the scope of this work, the right diagram for $d=4$ SAdS boundaries predicts that we will find fat funnels to dominate at small $R$ and that either fat or uniform funnels will dominate at large $R$.

\subsection{BTZ droplets and funnels: a brief review}
\label{BTZ}

It was argued in \cite{Hubeny:2009ru} that the most interesting droplet and funnel phases for BTZ boundary metrics could be found analytically due to an at-first-sight surprising SO(2,1) conformal symmetry of the boundary spacetimes.  This constitutes a symmetry of the boundary conditions which corresponds to a bulk isometry of any phase in which it is not spontaneously broken.    The phases preserving this symmetry can then be mapped via double Wick rotation to static spherically-symmetric solutions classified by Birkhoff's theorem.  In other words, any phase transition preserving this symmetry can be mapped to the familiar Hawking-Page transition between thermal AdS${}_4$ and Schwarzschild-AdS${}_4$.

To see this symmetry, we begin with the non-rotating\footnote{The rotating case was also discussed in \cite{Hubeny:2009rc}.} BTZ metric \cite{Banados:1992wn,Banados:1992gq}
\begin{equation}
\mathrm{d}s^2_{BTZ} =  \frac{r^2\ell_3^2}{R^4} \left[ - \left( 1 - \frac{R^2}{r^2}\right) \frac{R^4 \mathrm{d}t^2}{\ell_3^4} + \frac{R^4}{r^4} \frac{\mathrm{d}r^2}{1 - R^2/r^2} +  \frac{R^4}{\ell_3^2} \mathrm{d}\phi^2 \right],
\end{equation}
where we have pulled out an overall factor of $\frac{r^2}{\ell_3^2}$ relative to the usual presentation.  Introducing $\eta = R^2 t/\ell_3^2$, $\tau = R^2 \phi/\ell_3^2$, and  $\sin \theta = R/r$ this becomes
\begin{equation}
\label{dS2}
\mathrm{d}s^2_{BTZ} = \frac{\ell_3^2}{R^2 \sin^2 \theta} \left(-\cos^2 \theta\,\mathrm{d}\eta^2 + \mathrm{d} \theta^2 +  \mathrm{d}\tau^2 \right).
\end{equation}
Note that the factor in square brackets is just the $\theta \ge 0$ half of the static patch of dS${}_2 \times S^1$.  The full static patch is then obtained by gluing together two copies of the BTZ metric as shown in figure \ref{fig:bndy}.  Including the region behind the BTZ horizons leads to global dS${}_2 \times S^1$, as is clear from the fact that this is the maximal analytic continuation preserving periodicity of $\tau.$

\begin{figure}
\centerline{
\includegraphics[width=0.6\textwidth]{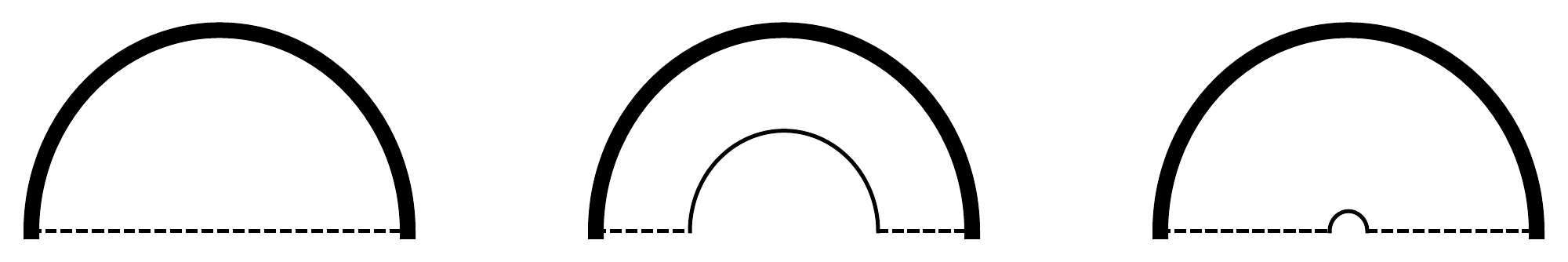}
}
\caption{Surfaces of constant Killing time and constant azimuthal angle for global AdS${}_4$ (left) and the large (center) and small (right) Schwarzschild-AdS${}_4$ black holes.  The upper solid semi-circle is the AdS boundary.  When it exists, the smaller solid semi-circle is the horizon.  Dashed lines are the azimuthal rotation axes.  Horizons and rotation axes are exchanged under double Wick rotations. Interpreting solid lines as rotation axes and dashed lines as horizons, these figures thus also depict static slices of spacetimes with BTZ boundary black holes corresponding to  the funnel (left) and short (center) and long (right) droplet phases discussed in the text.
}
\label{fig:HP}
\end{figure}
It is the SO(2,1) symmetry of global dS${}_2$ that allows analytic control.  Wick rotating dS${}_2$ to $S^2$, the boundary becomes $S^2 \times S^1$, which is just the thermal $d=3$ Euclidean ESU.   Applying the same operations to any bulk solution invariant under this SO(2,1) $\times$ U(1) symmetry gives a spherically-symmetric static solution asymptotic to empty (global) AdS${}_4$.  So in the absence of spontaneous symmetry breaking the phases we seek can be obtained by double Wick rotation of empty (thermal) AdS${}_4$ together with the large and small AdS-Schwarzschild black holes.

To understand the relation of these three geometries to funnels and droplets, consider a co-dimension 2 surface in each spacetime at some constant Killing time and constant azimuthal angle.  As shown in figure \ref{fig:HP}, such surfaces terminate on horizons (solid lines) and at the azimuthal rotation axis (dashed lines).  Double Wick rotation exchanges horizons with rotation axes, so one obtains spacetimes with BTZ boundaries where the solid line describes a rotation axis and the dotted lines describe horizons.  The solution obtained from AdS${}_4$ is a black funnel; it has no rotation axis and a horizon that runs from one side of the ESU${}_3$ boundary to the other.  Explicit calculation shows that it is in fact the BTZ black string of \cite{Emparan:1999fd}.  In contrast, double Wick rotation of the black hole spacetimes yields a rotation and two disconnected components to the horizon.  These are the desired droplet solutions, which we call short and long based on the distance the horizon penetrates into the bulk.

Since double Wick rotation does not change the Euclidean action, the details of the phase transition are equivalent to those of Hawking-Page \cite{Hawking:1982dh}.    But the mapping of parameters is non-trivial.  In particular, the high-temperature behavior of the Hawking-Page transition maps to physics of low-temperature BTZ boundary black holes.  The funnel phase exists for all temperatures as a local minimum of free energy.  It dominates the canonical ensmble below at $T_{BTZ} = \frac{1}{4 \pi \ell_3}$.  In contrast, the droplet phases exist only for $T_{BTZ} \le \frac{1}{2 \pi \sqrt{3} \ell_3}$.  The short droplet phase locally minimizes the free energy and dominates below the transition temperature.  The long droplet phase locally maximizes free energy and can be interpreted as mediating the transition between the other two phases.   It is natural to expect a similar structure for droplets in higher dimensions, though funnels become more complicated as we discuss below.

\section{Uniform and non-uniform phases of global black funnels}
\label{pert}

As mentioned above, further insight into the substructure of funnel phases can be obtained by considering the Gregory-Laflamme instability. After providing some general discussion below, section \ref{pert calcs} proceeds to perturbative calculations that generalize and extend the work of \cite{Hirayama:2001bi,Chamblin:2004vr,Gregory:2008br}.

Specializing to boundary spacetimes that describe global Schwarzschild-AdS${}_d$ (or BTZ) black holes, we may use the observation of \cite{Hubeny:2009rc} that the global AdS-Schwarzschild (or BTZ) string solutions provide exact funnel geometries with the desired boundary conditions for all $R/\ell_d$. The relevant bulk metrics are
\begin{equation}
\label{AdSSS}
\mathrm{d}s^2 = \frac{\ell^2_{d+1}}{\sin^2 z} \left(\mathrm{d}z^2 + \ell^{-2}_d \mathrm{d}s^2_{d}  \right)
\end{equation}
where $\ell_{d+1}, \ell_d$ are the bulk and boundary AdS length scales.  Here $z \in [0,\pi]$ and $ds^2_d$ is a metric for (global) Schwarzschild-AdS${}_d$ (BTZ):
\begin{equation}
\mathrm{d}s_d^2 = -f(r) \mathrm{d}t^2 + \frac{\mathrm{d}r^2}{f(r)} + r^2 \mathrm{d}\Omega_{d-2}^2,
\label{eq:schwarzschildAdS}
\end{equation}
with $d\Omega^2_{d-2}$ the metric on the unit $S^{d-2}$, $f = \frac{r^2}{\ell_d^2} + 1 - \frac{2M}{r^{d-3}}$ for $d \ge 4$, and $f =  \frac{r^2}{\ell_d^2} - 2M$ for $d=3$. The horizon size $R$ is the unique positive real root of the equation $f(r) = 0$. The BTZ solution was obtained in \cite{Emparan:1999fd} as a special case of the AdS C-metric while the $d \ge 4$ solutions were studied in \cite{Hirayama:2001bi}.

Such solutions may be constructed by first writing AdS${}_{d+1}$ in terms of AdS${}_d$ slices and then replacing each slice with Schwarzschild AdS${}_d$ (BTZ) of the same mass $M$, or equivalently with the same value of $R/\ell_d$.  Indeed \eqref{AdSSS} solves the vacuum bulk Einstein equations with cosmological constant so long as $\mathrm{d}s_d^2$ is an Einstein metric with Ricci scalar set in the usual way by $\ell_d^2$. The construction thus extends to the rotating case, though for simplicity we fix all angular momenta to zero.  The BTZ version was used in \cite{Fischetti:2012vt} as a basis for constructing funnels with heat flow.

It is convenient to use \eqref{AdSSS} to define boundary conditions.  For each $R/\ell_d$, we seek bulk metrics with the same leading-order behavior at $z \rightarrow 0$.  The ESU frame described in section \ref{intro} is defined by introducing  $\tilde z =  \frac{{\rm \ell_d\,sin} z}{\sqrt{r^2+{\ell_d^2} }}$ and rewriting \eqref{AdSSS} in terms of $\tilde z$ and $\widetilde{\mathrm{d}s}_d^2 =\frac{\ell_d^2\,\mathrm{d}s_d^2}{r^2 +\ell_d^2}$, where this $\widetilde{\mathrm{d}s}^2_d$ is indeed the ESU metric for $M \rightarrow 0$.   The two boundary metrics at $z = 0, \pi$ then combine to form a single boundary metric at $\tilde z =0$.  Since $r$ diverges at the ESU equator, it is natural to replace $r$ by $\rho = \frac{{\rm sign} \  ({\rm sin} z) }{r}$ so that the $\mathbb{Z}_2$ symmetry $z \rightarrow -z$ acts on the boundary as $\rho \rightarrow - \rho.$  For odd $d$ the metric is manifestly smooth, though for even $d$ it contains $M |\rho|^{d-1}$ and is only $C^{d-2}$.

While we will seek and find smooth bulk metrics in all cases, an interesting effect of the non-smooth boundaries for even $d$ is that the energies of our solutions will diverge in the this ESU frame.  As we will discuss in detail in section \ref{sec:diag}, this divergence is clearly associated with $\rho=0$ where the two black hole solutions are patched together and not with the AdS${}_d$ black hole horizons.  In the particular case $d=4$ studied below, the boundary metric is still $C^2$.  While it may at first seem surprising that a quantum field theory should have divergent energy simply because the spacetime on which it lives fails to be $C^3$, for $d=4$ this is in fact a natural consequence of both power counting and the conformal anomaly.  Indeed, it is well known that the conformal anomaly contributes terms to the stress tensor involving two derivatives of the Ricci scalar which can manifestly diverge when the metric fails to be $C^3$. Nevertheless, we will see that energy {\it differences} remain finite between distinct solutions sharing these boundary conditions.

Returning to \eqref{AdSSS}, inspection of the metric shows that $\partial_z$ is a conformal Killing field. We therefore refer to this solution as the uniform black funnel.  However, for $R/ \ell_d \ll 1$ and $d \ge 4$, the spacetime deep in the bulk near $z = \pi/2$ approximates that of the $\Lambda =0$ Schwarzschild black string.  As noted in \cite{Hirayama:2001bi,Chamblin:2004vr,Gregory:2008br}, this should result in an instability like that found by Gregory and Laflamme \cite{Gregory:1993vy}. We thus also expect to find non-uniform black funnels, analogous to the non-uniform black strings of \cite{Gregory:1993vy,Gubser:2001ac,Sorkin:2004qq,Wiseman:2002zc}, by following the zero-mode from the onset of the instability.

This zero mode fattens the funnel in some places and thins it in others.  Two distinct non-uniform solutions may be found by following the zero-mode with either sign.  For otherwise translationally invariant strings these solution are equivalent:  thinning the string in one place is equivalent to fattening it in another.  But our uniform funnels admit only a conformal Killing field.  In particular, the ${\mathbb Z}_2$ symmetry $z \rightarrow \pi - z$ gives a preferred middle ($z = \pi/2$) at which to compare the girth of the three funnels.  We therefore refer to the non-uniform phases as fat and thin, based on their sizes relative to the unform funnel at this point.

As we discuss below, following the zero-mode any finite distance will require us to deform the value of $R/\ell_d$ away from the value $R_{\mathrm{onset}}/\ell_d$ where the zero mode arises. Since our fat and thin funnels are distinct, following the zero mode in one direction should (at least initially) increase $R/\ell_d$ while following it in the other direction should (at least initially) decrease $R/\ell_d$.  One may thus wonder whether fat and thin funnels ever exist for the same value of $R/\ell_d$.  However, the Hawking-Page transition discussion in \ref{intro} suggests that both indeed exist at sufficiently small $R/\ell_d$.

This can be the case only if the branch of solutions (fat or thin) that moves from $R_{\mathrm{onset}}$ toward larger $R$ eventually turns around at some $R_{\mathrm{turn}}$ and returns to small $R$.  Such behavior is natural, as it agrees with that of the Hawking-Page transition where both big and small branches meet at the nucleation temperature $T_{\mathrm{nucl}} =\sqrt{(d-1)(d-3)}/(2\pi \ell_d)$, below which no black holes exist.  Phases of Kaluza-Klein black holes are also well-known to display similar behavior
\cite{Sorkin:2003ka,Kudoh:2003ki,Kudoh:2004hs,Kol:2005vy,Headrick:2009pv,Kalisch:2015via,Kalisch:2016fkm,Kalisch:2017bin}.

It is natural to suppose that our funnel phase diagram near $R_{\mathrm{turn}}$ resembles that of Hawking-Page \cite{Hawking:1982dh} near $T_{\mathrm{nucl}}$.  Thus the free energy $F$ should decrease with increasing thickness of the funnel and the uniform funnels should dominate.  We may also expect the behavior near $R_{\mathrm{onset}}$ to resemble that found in \cite{Gubser:2001ac,Sorkin:2004qq} so that the non-uniform funnels near $R_{\mathrm{onset}}$ have $F > F_{\mathrm{uniform}}$ for small dimension $d$ but $F < F_{\mathrm{uniform}}$ when $d$ is large.  In \cite{Sorkin:2004qq} the transition\footnote{For the case studied in \cite{Sorkin:2004qq}, $\Delta E$, $\Delta S$, and $\Delta F$ all change sign at the same value of $d$.} occurs between bulk dimensions $d+1=13$ and $d+1=14$. However, the perturbative analysis of uniform SAdS$_d$ funnels in section \ref{pert calcs} below gives $F < F_{\mathrm{uniform}}$ already in $d=4$, the lowest dimension where non-uniform funnels arise!

Finally, we expect the thin funnel phase to merge with some droplet phase in much the same way that thin non-uniform black string phases merge with phases describing Kaluza-Klein black holes in \cite{Kol:2002xz,Sorkin:2003ka,Kudoh:2003ki,Kudoh:2004hs,Kol:2005vy,Headrick:2009pv,Kalisch:2015via,Kalisch:2016fkm,Kalisch:2017bin}.  This expectation will be realized in section \ref{sec:results}.

\subsection{Perturbations of Uniform Black Funnels}
\label{pert calcs}

Having established our general expectations above, we now proceed to calculations. Let $\bar{g}$ represent our background metric, and $h$ its infinitesimal perturbation. Then, in the traceless transverse gauge where the metric perturbation satisfies
\begin{equation}
\bar{\nabla}^ah_{ab}=0,\quad\text{and} \quad h=0\,,
\end{equation}
the linearized Einstein equations take the form
\begin{equation}
\Delta_L h_{ab} \equiv \bar{\nabla}_c \bar{\nabla}^c h_{ab}+2 \bar{R}_{acbd}h^{cd} = 0\,,
\label{eq:Lichne}
\end{equation}
where over-barred quantities are computed using $\bar{g}$ and $h = h_{ab} \bar{g}^{ab}$.

The perturbative calculations are best understood if we take as the background the line element (\ref{AdSSS}), with $\mathrm{d}s^2_{d}$ as in Eq.~(\ref{eq:schwarzschildAdS}). For the case at hand, $\mathrm{d}s^2_{d}$ is an Einstein manifold, so we can decompose our perturbations according to how they transform under diffeomorphisms on $\mathrm{d}s^2_{d}$. The case that we are most interested in corresponds to tensor perturbations, which take the following particular form
\begin{equation}
h_{z\,a}=0,\quad h_{\mu\nu}=Y(z) \hat{h}_{\mu\nu}\,,
\label{eq:perturbansatz}
\end{equation}
where Greek indices indicate boundary directions and
\begin{equation}
\hat{\nabla}^\mu \hat{h}_{\mu\nu}=0\quad\text{and}\quad\hat{g}^{\mu\nu}\hat{h}_{\mu\nu}=0\,.
\label{eq:TT}
\end{equation}
Here, hatted quantities are computed with respect to $\hat{g}$, the metric on $\mathrm{d}s^2_{d}$. We are further interested in metric perturbations $\hat{h}$ that preserve spherical symmetry and do not depend on time\footnote{Note that we are searching for the zero mode, corresponding to a static perturbations.}, so that $\hat{h}$ is given by
\begin{equation}
\hat{h}_{\mu\nu}d x^\mu dx^\nu = a(r)f(r)dt^2+\frac{b(r)dr^2}{f(r)}+c(r) r^2d\Omega_{d-2}\,,
\end{equation}
where the factors of $f(r)$ in $\hat{h}_{tt}$ and $\hat{h}_{rr}$ and the factor of $r^2$ multiplying the $d-2$ sphere, were introduced for later convenience. Remarkably, the gauge conditions (\ref{eq:TT}) turn out to be algebraic in $a(r)$ and $c(r)$, which means they can be readily solved with respect to $b(r)$ and $b^\prime(r)$:
\begin{equation}
a(r)=\frac{2f(r)[d\,b(r)+r b^\prime(r)]}{2f(r)-r f^\prime(r)}-b(r)\,,\quad\text{and}\quad c(r)=\frac{r[d\,b(r)f^\prime(r)+2f(r)b^\prime(r)]}{(d-2)[2f(r)-r f^\prime(r)]}+b(r)\,,
\label{eq:gaugeconcrete}
\end{equation}
where ${}^{\prime}$ indicates derivative with respect to $r$.

We are now ready to determine the final equations resulting from tensor perturbations. First, we input the gauge conditions (\ref{eq:gaugeconcrete}) and the ansatz (\ref{eq:perturbansatz}) into the perturbed Einstein equations (\ref{eq:Lichne}). This will give one second and two third order equations in $r$ and $z$. This is not a surprise, since the gauge conditions are first order in $r$. The $rr$ component of $\Delta_L h$ is second order in both $z$ and $r$, and can be shown to solve the remaining two third order equations. Furthermore, as expected, we get two decoupled equations for $Y$ and $b$, with a separation constant $K$. These are:
\begin{subequations}
\begin{equation}
\ddot{Y}(z)-(d-5)\cot z\,\dot{Y}(z)-2(d-2)\cot^2z\,Y(z)+(K-2)Y(z)=0\,,
\label{eq:sepY}
\end{equation}
and
\begin{multline}
-f\,b''+\frac{2\,r^2(f\,f^{\prime\prime}-{f^\prime}^2)-r(d-2)f\,f^\prime+2\,d\,f^2}{r(r f^\prime-2 f)}\,b'
\\
+\frac{r^2\,f^{\prime}f^{\prime\prime}+r[2(d-1)f f^{\prime\prime}-(d+2){f^\prime}^2]+4f\,f^{\prime}}{r(r f^\prime-2 f)}b+\frac{K}{\ell^2_d}b=0\,,
\label{eq:sepb}
\end{multline}
\end{subequations}
where $\dot{}$ indicates derivative with respect to $z$. The equation for $b$ has appeared, in a different but related context, in \cite{Prestidge:1999uq}, where the negative mode of the Euclidean partition function of the Schwarzschild-AdS black hole was studied. There, $K$ was identified as the Euclidean negative mode. The fact that we get the same equation is not a surprise, since we expect the local thermodynamic stability of the Schwarzschild-AdS black hole to be related to the dynamical stability of the corresponding uniform funnel. This plays a central role in the Gubser-Mitra conjecture \cite{Gubser:2000ec,Gubser:2000mm}.

It turns our that Eq.~(\ref{eq:sepY}) has a simple analytic solution in terms of Hypergeometric functions of the second kind. Here we choose the solution that is automatically regular at one of the two singular points, $z=\pi$:
\begin{equation}
Y(z)={}_2F_1\left(\frac{d+1}{2}-\sqrt{\frac{(d-1)^2}{4}+K},\frac{d+1}{2}+\sqrt{\frac{(d-1)^2}{4}+K},1+\frac{d}{2},\cos^2\left(\frac{z}{2}\right)\right)\,\sin^{d-2}z\,.
\end{equation}
The quantization of $K$ comes from demanding a normalizable solution at $z=0$, which yields
\begin{equation}
K=(d+p)(p+1)\,,\quad \text{for}\quad p=0,1,2,\ldots\,.
\label{eq:quanti}
\end{equation}
We are thus left to solve Eq.~(\ref{eq:sepb}). Note that $\ell_{d+1}$ has completely decoupled from the problem, in particular, $f$ only depends on $r$, $M$ and $\ell_d$. For the sake of presentation, it is useful to parametrize our solution by the radius of the boundary black holes, normalized to $\ell_d$. This can be easily done by noting that on the horizon, located at $r=R$, $f(R)=0$. This allow us to rewrite $M$ as a function of $R/\ell_d$. Furthermore, we introduce the following compact coordinate:
\begin{equation}
\tilde{x}\equiv 1-\frac{R}{r},
\end{equation}
that takes values in the unit interval, \emph{i.e.} $\tilde{x}\in (0,1)$, being $1$ at asymptotic infinity and $0$ at the horizon. Now we also note that:
\begin{equation}
f(r) = \frac{1}{(1-x)^2}\left[-\left(\frac{R^2}{\ell^2_d}+1\right) (1-x)^{d-1}+\frac{R^2}{\ell^2_d}+(1-x)^2\right]\,,
\end{equation}
meaning that if we express $f(r)$ in term of $x$, it depends on
\begin{equation}
\rho\equiv \frac{R^2}{\ell^2_d}
\end{equation}
only. We are now ready to study the boundary conditions of $b$. In order to solve for $b$, we first need to investigate its boundary conditions. At the horizon, there are two possible solutions
\begin{equation}
b(x)=C_1+\frac{C_2}{x},
\end{equation}
and regularity demands $C_2=0$. On the other hand, close to $x=1$, we find the following two possible behaviors
\begin{equation}
b(x)=A_1 (1-x)^{1-p}+A_2 (1-x)^{d+p+2}\,.
\end{equation}
Normalizability at $x\to1$, requires $A_1=0$. Note that for $p=0$ one might wonder whether $A_1\neq0$ is an allowable solution as well. However, one can check that, by using Eq.~(\ref{eq:gaugeconcrete}), this would correspond to a divergent $a(r)$.

Now that we have unravelled which boundary conditions we want to consider, we proceed to the actual numerical method we have used. First, we introduce a new function, $q(x)$, that will actually be used in the numerics. This function is defined as
\begin{equation}
b(x) = (1-x)^{d+p+2}q(x)\,.
\end{equation}

The equation for $b$, namely Eq.~(\ref{eq:sepb}), is secretly a quadratic St\"urm-Liouville equation in $q(x)$, with quadratic eigenvalue $\rho$. It takes the following schematic form
\begin{equation}
\mathcal{L}^{(0)}\,q-\rho\,\mathcal{L}^{(1)}\,q-\rho^2\,\mathcal{L}^{(2)}\,q=0\,,
\end{equation}
where each of $\mathcal{L}$'s is a second order differential operator in $x$, that is $\rho$ independent. Their explicit form can be found in appendix \ref{ap:1}. We can now use standard methods to solve this linear boundary value problem, see for instance \cite{Dias:2015nua}.

\begin{figure}
\centerline{
\includegraphics[width=0.5\textwidth]{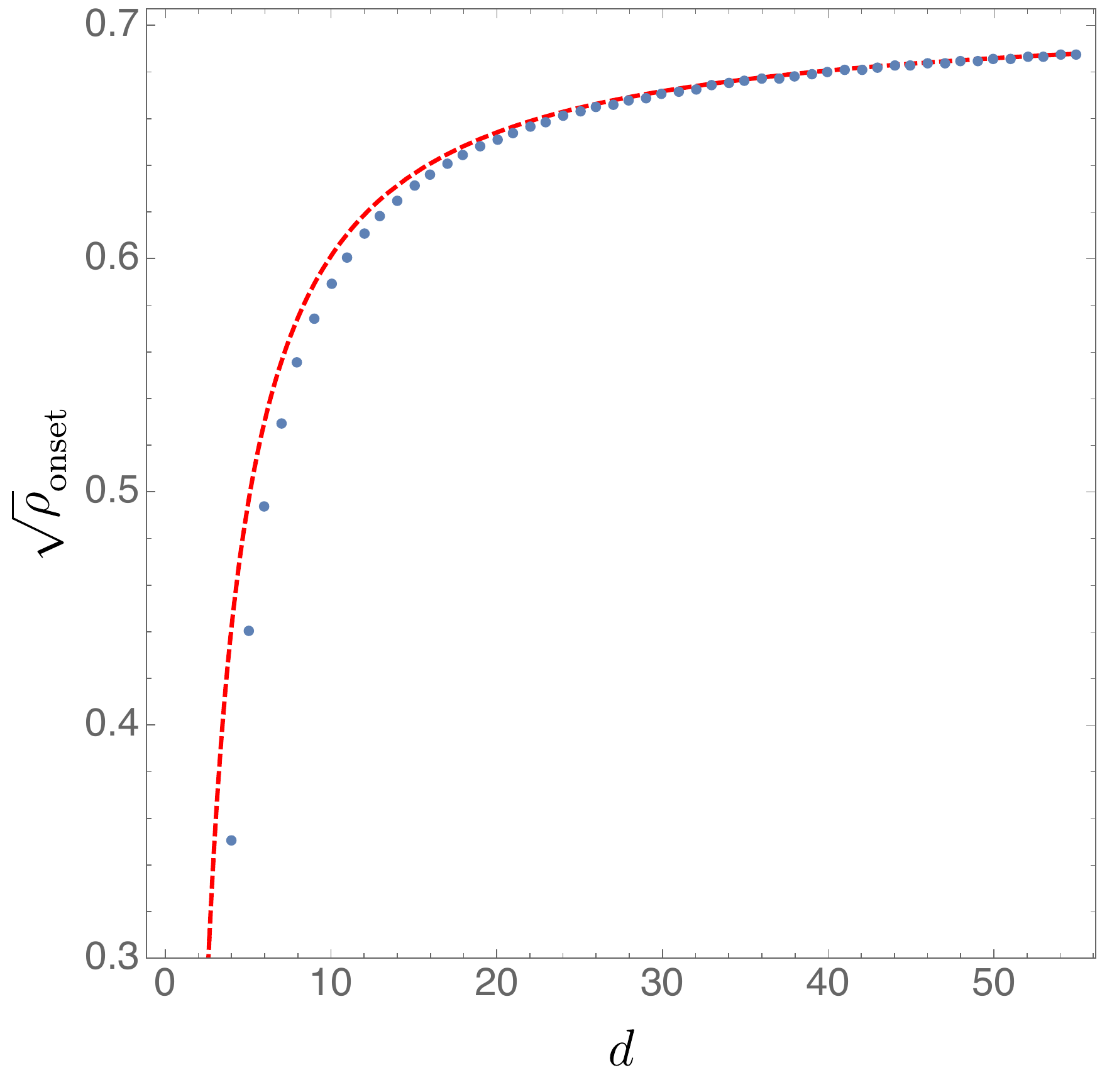}
}
\caption{$R_{\mathrm{onset}}/\ell_d\equiv \sqrt{\rho}_{\mathrm{onset}}$ as a function of $d$: the blue disks represent our exact numerical data, and the dashed red line the analytic expression (\ref{eq:larged}), valid at large $d$.}
\label{fig:criticalR}
\end{figure}

Numerical results for $R_{\mathrm{onset}}/\ell_d$ are plotted as a function of $d$ in Fig.~\ref{fig:criticalR}. As expected, $R_{\mathrm{onset}}/\ell_d$ is a monotonic function of $d$. This is exactly what we expect, since for high $d$ it should be easier to render the uniform funnel unstable. Also, we find that there is no instability for $d=3$, in contrast to the expectations of \cite{Emparan:1999fd}.

We can repeat \emph{mutatis mutandis} the $1/d$ expansion of \cite{Asnin:2007rw} to find a closed form expression for $\sqrt{\rho}_{\mathrm{onset}}$ at large $d$
\begin{equation}
\sqrt{\rho}_{\mathrm{onset}}=\frac{1}{\sqrt{2}}\left[1-\frac{3}{2\,d}+\mathcal{O}(d^{-2})\right]\,.
\label{eq:larged}
\end{equation}
Fig.~\ref{fig:criticalR} also plots $\sqrt{\rho}_{\mathrm{onset}}$ as given by Eq.~(\ref{eq:larged}) and finds excellent agreement at large $d$ (for instance, for $d=55$ we find that the relative error between the exact numerical value and our analytic expression is smaller than $6\times 10^{-2} \%$). Note also that our onset curve shows that $R_{\mathrm{onset}}$ is always smaller than $R_{\mathrm{nucl}}$ (where large and small AdS black holes meet), regardless of the number of dimensions.

We were also able to go take our perturbative approach beyond first order. The scheme is very similar to that used in \cite{Dias:2011ss}, except that we cannot use the gauge invariant formalism of \cite{Kodama:2003jz}. As such we need to start with a gauge choice, which we detail in the next section.

We start assuming that a general metric, sufficiently close to $\bar{g}_{ab}$, admits an expansion of the form
\begin{equation}
g_{ab} = \bar{g}_{ab}+\sum_{i=1}^{+\infty} \varepsilon^i H^{(i)}_{ab}\,.
\end{equation}
At each order in perturbation theory, Einstein equations always take the same form, namely
\begin{equation}
\tilde{\Delta}_L H^{(i)}_{ab}=T^{(i)}_{ab}\,,
\end{equation}
where $\tilde{\Delta}_L$ is a second order operator that only depends on $\bar{g}$ and $T^{(i)}$ is an effective stress energy tensor that generically depends on all of the $H^{(j<i)}$ and their derivatives $i-$the derivatives. Because $\tilde{\Delta}_L$ only depends on the background, it admits a natural decomposition in terms of perturbation with definite transformation properties under diffeomorphisms on $\mathrm{d}s^2_d$. These can be tensor metric perturbations, vector metric perturbations or scalar metric perturbations. Schematically,
\begin{equation}
H^{(i)}_{ab} = \sum_{k}\left[h^{(S),(i),(k)}_{ab}+h^{(T),(i),(k)}_{ab}+h^{(V),(i),(k)}_{ab}\right]\,,
\end{equation}
where $k$ represents the quantum numbers of each of these perturbations. For instance, for tensor perturbations, $k$ would be $p$ in Eq.~(\ref{eq:quanti}).The idea is to start at linear order with one of these modes, for instance the tensor perturbation discussed above, and compute $T^{(2)}_{ab}$. We then decompose $T^{(2)}_{ab}$ as a sum of tensors, vectors or scalars perturbations and compute $h^{(2)}_{ab}$. At second order one finds that there is a tensor metric perturbation, with the same quantum numbers as the one you started with, that cannot be made regular. It turns out, this singularity can be removed by promoting $\rho$ to be $\varepsilon$ dependent:
\begin{equation}
\rho = \rho^{(0)}+\sum_{i=1}^{+\infty} \varepsilon^{i} \rho^{(i)}\,,
\end{equation}
where $\rho^{(0)}$ was determined from the eigenvalue problem detailed above. The singularity at second order can be readily removed by an appropriate choice of $\rho^{(1)}$.

Doing so and computing $\Delta F =  F- F_{\mathrm{uniform}}$ yields the
2nd order perturbative result \begin{equation}
\ell_{4}\Delta F/N^2 \approx -10.6571 (T\,\ell_5-0.312617)^2\,
\label{eq:pert}
\end{equation}
in terms of the parameter
\begin{equation}
\label{eq:N2}
N^2=\frac{\pi}{2}\frac{\ell_5^3}{G_5}\,.
\end{equation}
We use this parameter for convenience, but it also describes the rank of the gauge group of the dual super Yang-Mills theory in an AdS/CFT context \cite{Maldacena:1997re}.  As noted above, the $d=4$ result \eqref{eq:pert} is a surprise when compared with the analogous study of the original Gregory-Laflamme instability in \cite{Sorkin:2004qq} which found $\Delta F < 0$ only for bulk dimensions $d+2 \ge 14$.  We now turn to constructing and analyzing the corresponding non-perturbative solutions numerically.   In section \ref{sec:results} we will present results that fit \eqref{eq:pert} well for $T\,\ell_5 \approx 0.312617$.

\section{Numerical Procedure and Framework}
\label{numerics}

Having set our expectations with analytic arguments, we now turn to numerics to explore details of the phase diagrams. We present new results only for $d=4$, though for $d=3$ our code also reproduces the analytic results of \cite{Hubeny:2009rc}. In addition, for $d=3$ our code also allows for boundary black holes that are not BTZ.

We will first introduce the general numerical technique that allowed us to determine the phase diagram of black funnels and droplets. This technique was first introduced in \cite{Adam:2011dn}, discussed in great detail in \cite{Figueras:2011va} and reviewed in \cite{Wiseman:2011by,Dias:2015nua}. The idea is to solve a set of PDE's that are manifestly Elliptic, and whose solutions coincide with solutions of the Einstein equations in a certain gauge.

We deform Einstein's equations:
\begin{equation}
G_{ab} \equiv R_{ab}-\frac{d}{\ell_{d+1}^2}g_{ab}=0\,,
\label{eq:einstein}
\end{equation}
by adding the following new term
\begin{equation}
G^{H}_{ab} \equiv G_{ab}-\nabla_{(a}\xi_{b)}=0,
\label{eq:einsteindeturck}
\end{equation}
where $\xi^a = g^{cd}[\Gamma^a_{cd}(g)-\bar{\Gamma}^a_{cd}(\bar{g})]$ and $\bar{\Gamma}(\bar{g})$ is the Levi-Civita connection associated with a reference metric $\bar{g}$. It is easy to show that any solution to $G_{ab}=0$ with $\xi=0$ is a solution to $G^{H}_{ab}=0$. However, the converse is not necessarily true. In certain circumstances one can show that solutions with $\xi\neq 0$, coined Ricci solitons, cannot exist \cite{Figueras:2011va}. The line elements discussed in this manuscript are exactly in that class. In particular, in order to ensure that $\xi$ is everywhere zero, we need to ensure that all components of $\xi$ are zero at any asymptotic end, and that the extrinsic curvature at fictitious boundaries, such as horizons, is zero. One can show that these conditions are only sufficient, and that $\xi$ can be made zero for more general boundary conditions \cite{Figueras:2011va}.

A choice of reference metric is equivalent to a gauge choice, which in the DeTurck formalism is given by a natural generalization of the Harmonic gauge $\xi=0\Leftrightarrow \triangle x^a = g^{cd}\bar{\Gamma}^a_{cd}(\bar{g})$. As a result, some reference metrics are more amenable to numerics than others. Our criteria for choosing the reference metric is simple: we demand it has the same axis and horizon locations as the metric we seek to find, and satisfies the same Dirichlet boundary conditions as $g$. We have used a standard pseudospectral collocation approximation on Chebyshev-Gauss-Lobatto points and solved the resulting non-linear algebraic equations using a  Newton-Raphson method. This discretization is well know to have exponential convergence, so long as all functions are analytic in their integration domain. As we shall see below, this is not the case for the Einstein DeTurck equation. This issue is particularly relevant when reading asymptotic quantities such as the total energy of a given solution \footnote{In appendix \ref{ap:2} we study convergence of our solutions in the continuum limit, and find evidence for power law convergence.}.

Implementing this procedure in particular cases requires a choice of metric ansatz and boundary conditions.  We describe these in turn for each of the situations we wish to study.  Some readers may wish to skip directly to our diagnostics in section \ref{sec:diag} or to section \ref{sec:results} which presents our numerical results.

\subsection{Ansatz for black funnels in $d = 4$}
The line element we use to describe thin and fat funnels takes the following form
\begin{multline}
\mathrm{d} s^2 = \frac{\ell^2_5}{1-y^2}\Bigg\{\frac{1}{(1-x^2)^2}\Bigg[-x^2\,g(x)\,A\,\mathrm{d}\tilde{t}^2+\frac{4\,\rho\,S_1\,}{g(x)}\left[\mathrm{d}x+(1-x^2)^2F\,\mathrm{d}y\right]^2
\\
+\rho\,S_2\mathrm{d}\Omega^2_2\Bigg]+\frac{B\mathrm{d}y^2}{1-y^2}\Bigg\}\,,
\label{eq:thinfat}
\end{multline}
where we recall that $\rho\equiv R^2/\ell^2_4$ and
\begin{equation}
g(x)=(1-x^2)^2 + (3-3\,x^2+x^4)\,\rho\,.
\label{eq:gf}
\end{equation}
$A$, $B$, $F$, $S_1$ and $S_2$ are function of $x$ and $y$ to be determined in our numerical procedure. Here, $(x,y)$ take values in the unit square, with $x=0$ being the funnel horizon, $y=1$ the conformal boundary $x=1$ the point infinitely far away from the black funnel, and $y=0$ the plane of symmetry that divides the funnel into two equal halves. Finally, for reference metric we take the line element above with $A=B=S_1=S_2=1$ and $F=0$.

\subsection{Boundary conditions at the horizon $x=0$}
The metric (\ref{eq:thinfat}) and the associated reference metric are regular at $x=0$, if and only if $A(0,y)=S_1(0,y)$. The best way to understand this is to work in the Euclidean section, setting $\tilde{t}=-i\tau$, and introducing the new coordinate
\begin{equation}
x \equiv \frac{\sqrt{g(0)}}{2\sqrt{\rho}}\eta\,,
\end{equation}
which brings the line element \ref{eq:thinfat} to the following form
\begin{multline}
d s^2 \approx \frac{\ell^2_{d+1}}{1-y^2}\Bigg\{\eta^2\,A(0,y)\left[\frac{g(0)\,d\tau}{2 \sqrt{\rho}}\right]^2+S_1(0,y)d\eta^2+\rho\,S_2(0,y)d\Omega^2_2+ \frac{4\sqrt{\rho}S_1(0,y)\,F(0,y)}{\sqrt{g(0)}} \mathrm{d}\eta\mathrm{d}y\\
+\frac{4\,\rho\,S_1(0,y)\,}{g(0)}F(0,y)^2\mathrm{d}y^2+\frac{B(0,y)\mathrm{d}y^2}{1-y^2}\Bigg\}\,,
\label{eq:thinfatx0}
\end{multline}
where we have expanded all functions around $\eta=0$. We recognise the first two terms as being flat space, if and only if $A(0,y)=S_1(0,y)$ and if $\tau$ gets a periodicity of $4\pi\sqrt{\rho}/g(0)$ and if $F(x,y)= x\tilde{F}(x,y)$ for a smooth function $\tilde{F}(x,y)$. This in turn implies that our funnels have a temperature, measured in units of $[\tilde{t}]^{-1}$, parametrized by $\rho$, and given by
\begin{equation}
T_H = \frac{g(0)}{4\pi\sqrt \rho}=\frac{1+3\,\rho}{4\pi\sqrt\rho}\,.
\end{equation}
Furthermore, the extrinsic curvature at $\eta = 0$ is zero, which is one of the conditions detailed in \cite{Figueras:2011va} for the nonexistence of DeTurck solitons. The remaining functions all have Neumann-type boundary conditions, which can be found by expanding the equations of motion close to $x=0$. To wit, we find
\begin{equation}
A(0,y)=S_1(0,y)\,,\quad F(0,y)= 0\,,\quad \text{and}\quad \left.\frac{\partial A}{\partial x}\right|_{x=0}=\left.\frac{\partial B}{\partial x}\right|_{x=0}=\left.\frac{\partial S_1}{\partial x}\right|_{x=0}=\left.\frac{\partial S_2}{\partial x}\right|_{x=0}=0\,.
\end{equation}

\subsection{Boundary conditions at the asymptotic end $x=1$}
In \cite{Figueras:2011va} it was shown that asymptotic ends require $\xi^a=0$ on all components. If we impose $A=B=S_1=S_2=1$ and $F=0$ that turns out to be the case.
\subsection{Boundary conditions at the reflection plane $y=0$}
The boundary conditions that we are interested at $y=0$ are those of a reflection plane. Equivalently, we want the extrinsic curvature on the induced hyperslice defined by $y=0$ to vanish. This can be easily achieve if we demand
\begin{equation}
\left.\partial_y A\right|_{y=0}=\left.\partial_y B\right|_{y=0}=\left.\partial_y S_1\right|_{y=0}=\left.\partial_y S_2\right|_{y=0}=F(x,0)=0\,.
\end{equation}
As alluded above, these boundary conditions also ensure that no DeTurck solitons exist.
\subsection{Boundary conditions at the conformal boundary $y=1$}
Here we shall not only discuss the relevant boundary conditions, but also how to extract the corresponding stress energy tensor. At the conformal boundary we impose
\begin{equation}
A(x,1)=B(x,1)=S_1(x,1)=S_2(x,1)=1\,\quad \text{and}\quad F(x,1)=0\,.
\end{equation}
This automatically ensures that both $\xi^x$ and $\xi^y$ are zero at the boundary, and thus that no DeTurck solitons exist in our spacetime.

One can also solve the equations off the conformal boundary up to any desired order. These take the following schematic form
\begin{subequations}
\begin{align}
&A(x,y)=1+ \delta(x)(1-y)^2+\alpha(x)(1-y)^{1+\sqrt{3}}+o[(1-y)^{1+\sqrt{3}}]
\\
&B(x,y)=1+\beta(x)(1-y)^{1+\sqrt{3}}+o[(1-y)^{1+\sqrt{3}}]
\\
&F(x,y)=\gamma(x)(1-y)^2+\frac{g(x)}{24\rho}\varphi(x)(1-y)^2 \log(1-y)\nonumber
\\
&\qquad\qquad\qquad+\frac{g(x)\left[\alpha '(x)-\beta '(x)\right]}{8 \left(1+\sqrt{3}\right) \rho }(1-y)^{1+\sqrt{3}}+o[(1-y)^{1+\sqrt{3}}]
\\
&S_1(x,y)=1+\epsilon(x)(1-y)^2+\alpha(x)(1-y)^{1+\sqrt{3}}+o[(1-y)^{1+\sqrt{3}}]
\\
&S_2(x,y)=1-\frac{\epsilon(x)+\delta(x)}{2}(1-y)^2+\alpha(x)(1-y)^{1+\sqrt{3}}+o[(1-y)^{1+\sqrt{3}}]
\end{align}
\label{eqs:irrfunnels}
\end{subequations}
where all higher order terms depend on the five unknown functions $\{\alpha(x),\beta(x),\gamma(x),\delta(x),\epsilon(x)\}$ and their derivatives along $x$ and
\begin{equation}
\varphi(x) \equiv \epsilon^\prime(x)+\frac{8 x}{1-x^2}\epsilon(x)-\frac{x g^\prime(x)+2 g(x)}{2 x g(x)}\left[\delta (x)-\epsilon(x)\right]\,.
\end{equation}
Note that one can easily extract $\delta(x)$ and $\epsilon(x)$ by taking two derivatives of $A$ and $S_1$ with respect to $y$, respectively, and evaluate them at the conformal boundary. All of these five functions are to be determined by requiring regularity in the bulk. It is not surprising that there are five such integration functions, since our PDE system is second order, and there are a total of five functions to solve for. We thus expect a generic expansion consistent with $10$ free functions, half of which should be killed by our choice of boundary conditions, giving a total of five free functions. Note also that we expect $\xi^\mu=0$ to emerge as the unique solution, but it is not a condition that one can see emerging locally by solving the Einstein DeTurck equation off the AdS boundary. One can, however, see what \emph{local} conditions do come out by imposing $\xi^\mu$ in the asymptotic expansion. As we shall see, these are related to the conservation of the holographic stress energy momentum tensor.

Requiring $\xi^\mu$ to be zero order by order in a $(1-y)$ expansion further demands
\begin{equation}
\alpha(x)=\frac{1}{4} \left(1-\sqrt{3}\right) \beta (x)\,,\quad \text{and}\quad \varphi(x)=0\,.
\label{eq:csilocal}
\end{equation}
The first of these conditions ensures that the non-analytic piece that populates our expansion is pure gauge, since it can be reabsorbed via a redefinition of $y$. The second condition is related to the conservation of the holographic stress energy tensor, which we will extract next.

In order to read off the stress energy tensor, we closely follow the procedure first outlined in \cite{deHaro:2000xn}. First we change to Fefferman-Graham coordinates and fix the conformal frame. Because we only know our functions numerically, we can only preform this coordinate change asymptotically. Up to $\mathcal{O}(z^5)$, we find that the relevant coordinate transformation is given by
\begin{equation}
\tilde{t}=\ell_4^{-1}\,t\,,\quad x=\chi+\mathcal{O}(z^{4+2\sqrt{3}}),\quad y = 1-\frac{z^2}{2\,\ell_4^2}+\frac{z^4}{8\,\ell_4^4}-\frac{z^6}{32 \ell_4^6}+\mathcal{O}(z^{4+2\sqrt{3}})\,,
\end{equation}
which induces the following conformal boundary metric
\begin{equation}
\mathrm{d}s^2_\partial = \frac{1}{(1-\chi^2)^2}\left[-\chi^2\,g(\chi)\,\mathrm{d}t^2+\frac{4R^2\,\mathrm{d} \chi^2}{g(\chi)}+R^2\,\mathrm{d}\Omega^2_2\right]\,.
\end{equation}
If we further define $r = R/(1-\chi^2)$, we recover the line element (\ref{eq:schwarzschildAdS}) with $d=4$. This confirms that our line element has as a boundary metric two copies of a Schwarzschild AdS$_4$ black hole.

The $\chi-$dependent expectation value for the holographic stress energy tensor induced by the above coordinate transformation reads:
\begin{subequations}
\begin{align}
& T_{t\phantom{t}}^{\phantom{t}t}=\frac{\ell^3_5}{16\pi G_5\,\ell_4^4}\left[\delta(\chi)-\frac{3}{4}\right]
\\
& T_{\chi\phantom{\chi}}^{\phantom{\chi}\chi}=\frac{\ell^3_5}{16\pi G_5\,\ell_4^4}\left[\epsilon(\chi)-\frac{3}{4}\right]
\\
& T_{\Omega_i\phantom{\Omega_j}}^{\phantom{\Omega_i}\Omega_j} =-\frac{1}{2}\left( T_{t\phantom{t}}^{\phantom{t}t}+ T_{\chi\phantom{\chi}}^{\phantom{\chi}\chi}+\frac{3 \ell _5^3}{16 \pi  G_5 \ell _4^4}\right)\delta_{\Omega_i\phantom{\Omega_j}}^{\phantom{\Omega_i}\Omega_j}
\end{align}
\label{eq:holographicstress}
\end{subequations}
where $\Omega_i$ denotes any coordinate on the two sphere. The constant offsets in each of the components corresponds to the stress energy tensor of the uniform funnel, and the fact that is constant has been subject of intense study in the literature \cite{Gregory:2008br}. It results solely from the conformal anomaly, and is absent in uniform funnels that live in $d+1$ even bulk spacetime dimensions. It is easy to check that the holographic stress energy tensor satisfies
\begin{equation}
 T_{\mu}^{\phantom{\mu}\mu}= -\frac{3 \ell _5^3}{16 G_5 \pi  \ell _4^4}\,,\qquad\text{and}\qquad \nabla^\mu T_{\mu\nu}=0
\end{equation}
with the latter condition being enforced via the last equation in Eq.~(\ref{eq:csilocal}), and the first reproducing the standard four-dimensional conformal anomaly.
\subsection{Ansatz for black droplets in $d = 4$}
The metric ansatz for the black droplets takes the following form
\begin{multline}
\mathrm{d}s^2 = \frac{\ell^2_5}{1-y^2}\Bigg\{-\frac{(1-x)^2\,G(x)\,A(x,y)\mathrm{d}\hat{t}^2}{\rho}+\frac{4\,S_1(x,y)}{G(x)}\left[\mathrm{d}x+F(x,y) \mathrm{d}y\right]^2\\
+S_2(x,y)y^2\mathrm{d}\Omega^2_2+\frac{B(x,y)\mathrm{d}y^2}{1-y^2}\Bigg\}\,,
\label{eq:dropletline}
\end{multline}
where
\begin{equation}
G(x)=x^2(2-x)^2 + \left[1 + x(2-x) + x^2(2-x)^2\right] \rho\,.
\end{equation}
Alike in the funnels case $A$, $B$, $F$, $S_1$ and $S_2$ are function of $x$ and $y$ to be determined in our numerical procedure and $(x,y)$ take values in the unit square. Here $x=1$ is the droplets horizon, $x=0$ is the plane of symmetry that divides the droplet into two halves, $y=0$ is the axis of symmetry and $y=1$ is the conformal boundary. For reference metric we take the line element above with $A=B=S_1=S_2=1$ and $F=0$.

\subsection{Boundary conditions at the horizon $x = 1$}
The boundary conditions at $x=1$ are very similar to those of the black funnels at $x=0$. In particular, the line element (\ref{eq:dropletline}) and its associate reference metric are only regular at $x=1$ if $A(1,y)=S_1(1,y)$ and $F(1,y)=0$. This can be best understood if we again consider a coordinate transformation of the form
\begin{equation}
x \equiv 1-\frac{\sqrt{G(1)}\eta}{2}\,,
\end{equation}
and expand the metric (\ref{eq:dropletline}) around $\eta = 0$. As before, we arrive at a temperature, measured in units of $[\hat{t}]^{-1}$, given by
\begin{equation}
T_H = \frac{G(1)}{4\pi \sqrt{\rho}}=\frac{1+3\rho}{4\pi \sqrt{\rho}}\,.
\end{equation}
We can also expand the line element (\ref{eq:dropletline}) close to $\eta = 0$, to find that the extrinsic curvature at the horizon vanishes, which is one of the conditions detailed in \cite{Figueras:2011va} for the nonexistence of DeTurck solitons. For the remaining variables we find:
\begin{equation}
A(1,y)=S_1(1,y)\,,\quad F(1,y)= 0\,,\quad \text{and}\quad \left.\frac{\partial A}{\partial x}\right|_{x=1}=\left.\frac{\partial B}{\partial x}\right|_{x=1}=\left.\frac{\partial S_1}{\partial x}\right|_{x=1}=\left.\frac{\partial S_2}{\partial x}\right|_{x=1}=0\,.
\end{equation}
\subsection{Boundary conditions at reflection plane $x = 0$}
The boundary conditions at the reflection plane are perhaps a bit more involved. Suppose that we define your functions for $x \ge 0$ but we wish to use the line element (\ref{eq:dropletline}) to define a $\mathbb{Z}_2$-symmetric metric that extends to $x < 0$.  Under what conditions is this extended metric smooth at $x=0$?

In general, the requirement can be stated as follows:  consider Gaussian normal coordinates around the $x=0$ hypersurface with $\lambda$ the ``normal" coordinate normalized to measure proper distance along geodesics that orthogonally intersect $x=0$.  We set $\lambda =0$ when $x=0$. Then the power series expansion of the spacetime metric in these coordinates should contain only even powers of $\lambda$, so long as we are away from the boundary.

One introduces Gaussian normal coordinates perturbatively in an expansion in $\lambda$. Schematically, they take the following form:
\begin{equation}
x= \sum_{i=1}^{+\infty} A_i(Y)\lambda^i, \quad \text{and}\quad y= Y+\sum_{i=1}^{+\infty} B_i(Y)\lambda^i\,.
\end{equation}
The $\{A_i,B_i\}$ coefficients can be determined by demanding $g_{\lambda\lambda} = 1$ and $g_{\lambda Y}=0$. Note that there are no other cross terms. We can now look at the expansions in $\lambda$ of the remaining metric components, and ask whether they are even in $\lambda$. If we impose the Einstein-DeTurck equations, together with the following boundary conditions at $x=0$
\begin{equation}
\left.\partial_x A(x,y)\right|_{x=0}=\left.\partial_x B(x,y)\right|_{x=0}=\left.\partial_x S_1(x,y)\right|_{x=0}=\left.\partial_x S_2(x,y)\right|_{x=0}=F(0,y)=0\,,
\end{equation}
that turns out to be the case. We went up to fifth order, and we are convinced this will happen to all orders. Note that that there are some rather complicated cancellations of the odd oder terms in $\lambda$ which only occur if the equations of motion are used. This of course also implies zero extrinsic curvature at $x=0$, and thus that no DeTurck solitons exist.
\subsection{Boundary conditions at the axis $y = 0$}
These boundary conditions are obtained demanding that the axis is smooth. In order for this to happen we need $S_2(x,0)=B(x,0)$ and $F(x,0)=0$, so that the last two terms in (\ref{eq:dropletline}) combine to form the origin of three-dimensional flat space, with $y$ playing the role of radial coordinate. The regularity of the remaining metric functions translates into Neumman boundary conditions of the following form
\begin{equation}
\left.\partial_y A(x,y)\right|_{y=0}=\left.\partial_y B(x,y)\right|_{y=0}=\left.\partial_y S_1(x,y)\right|_{y=0}=\left.\partial_y S_2(x,y)\right|_{y=0}=F(x,0)=0\,.
\end{equation}
\subsection{Boundary conditions at the conformal boundary $y = 1$}
In this section we will not only detail the relevant boundary conditions at $y=1$, but also how to extract the holographic stress energy tensor.

We choose the following set of boundary conditions
\begin{equation}
A(x,1)=B(x,1)=S_1(x,1)=S_2(x,1)=1\quad\text{and}\quad F(x,1)=0\,.
\end{equation}
These automatically ensure that the leading terms of $\xi^a= \mathcal{O}(1-y)$, without the use of the equations of motion, and once more are within the class of boundary conditions studied in \cite{Figueras:2011va}, for which DeTurck solutions can be ruled out.

One can solve the equations in an expansion off the conformal boundary, to any desired order, and in particular we find:
\begin{subequations}
\begin{align}
&A(x,y)=1-2x(2-x)(1+\rho)(1-y)+ \delta(x)(1-y)^2\nonumber
\\
&\qquad\qquad\qquad\qquad\qquad\qquad\qquad+\alpha(x)(1-y)^{1+\sqrt{3}}+o[(1-y)^{1+\sqrt{3}}]\,,
\\
&B(x,y)=1+x (2-x) [2-x (2-x) (1+\rho )] (1+\rho )(1-y)^2\nonumber
\\
&\qquad\qquad\qquad\qquad\qquad\qquad\qquad+\beta(x)(1-y)^{1+\sqrt{3}}+o[(1-y)^{1+\sqrt{3}}]\,,
\\
&F(x,y)=\frac{1}{4} (1-x) (1+\rho ) G(x)(1-y)+\gamma(x)(1-y)^2\nonumber
\\
&\qquad\qquad\qquad+\frac{G(x)}{24}\left[\varphi(x)-9 x (2-x) (1-x)(1+\rho )^2\right](1-y)^2 \log(1-y)\nonumber
\\
&\qquad\qquad\qquad+\frac{G(x)\left[\alpha '(x)-\beta '(x)\right]}{8 \left(1+\sqrt{3}\right)}(1-y)^{1+\sqrt{3}}+o[(1-y)^{1+\sqrt{3}}]\,,
\\
&S_1(x,y)=1+2x(2-x)(1+\rho)(1-y)+\epsilon(x)(1-y)^2\nonumber
\\
&\qquad\qquad\qquad\qquad\qquad\qquad\qquad+\alpha(x)(1-y)^{1+\sqrt{3}}+o[(1-y)^{1+\sqrt{3}}]\,,
\\
&S_2(x,y)=1+x(2-x)(1+\rho)(1-y)-\frac{1}{2} [\delta (x)+\epsilon (x)+x (2-x) \hat{\varphi}(x) (1+\rho )](1-y)^2\nonumber
\\
&\qquad\qquad\qquad\qquad\qquad\qquad\qquad+\alpha(x)(1-y)^{1+\sqrt{3}}+o[(1-y)^{1+\sqrt{3}}]\,,
\end{align}
\label{eqs:expansiony1}
\end{subequations}
where
\begin{equation}
\varphi(x)\equiv \epsilon^\prime(x)-\frac{(1-x) G^\prime(x)-2 G(x)}{2 G(x) (1-x)}\left[\delta (x)-\epsilon (x)\right]\qquad\text{and}\qquad \hat{\varphi}(x)=1-5 x (2-x) (1+\rho )\,.
\end{equation}

By inspecting the local conditions coming from solving $\xi^a=0$ order by order in a $(1-y)$ expansion, one further finds
\begin{equation}
\alpha(x)=\frac{1}{4} \left(1-\sqrt{3}\right) \beta (x)\,,\quad \text{and}\quad \varphi(x)=0\,.
\end{equation}
Just like for the funnels, the first condition ensures that the terms proportional to $(1-y)^{1+\sqrt{3}}$ are pure gauge, and the last enforces the conservation of the holographic stress energy tensor.

Alike for the black funnels, we now change to Fefferman-Graham coordinates and fix the conformal frame. This can only be done perturbatively off the boundary, since we only know our functions numerically. We want to work in the same conformal frame we did before, since we want to explicitly compare the stress energy tensors of both phase. This is accomplished by the following coordinate transformation
\begin{subequations}
\begin{align}
&\tilde{t}=\ell_4^{-1}\,t\,,\\
&x = 1-\chi -\frac{1}{4} \chi  \left(1-\chi ^2\right) g(\chi )\frac{z^2}{R^2}-\frac{1}{32} \chi  \left(1-\chi ^2\right) g(\chi ) \partial_{\chi} \left[\chi  \left(1-\chi ^2\right) g(\chi )\right]+\mathcal{O}(z^6)\\
&y = 1-\frac{1}{2} \left(1-\chi ^2\right)^2\frac{z^2}{R^2}+\frac{1}{4}\left(1-\chi ^2\right)^2 \left[\chi ^2 g(\chi )+\frac{1}{2} \left(1-\chi ^2\right)^2\right] \frac{z^4}{R^4}+\mathcal{O}(z^6)\,,
\end{align}
where
\begin{equation}
g(\chi)=G(1-\chi)
\end{equation}
was defined in Eq.~(\ref{eq:gf}).
\end{subequations}
This coordinate transformation induces the following conformal boundary metric
\begin{equation}
\mathrm{d}s^2_\partial = \frac{1}{(1-\chi^2)^2}\left[-\chi^2\,g(\chi)\,\mathrm{d}t^2+\frac{4R^2\,\mathrm{d} \chi^2}{g(\chi)}+R^2\,\mathrm{d}\Omega^2_2\right]\,.
\end{equation}
If we further define $r = R/(1-\chi^2)$, we recover the line element (\ref{eq:schwarzschildAdS}) with $d=4$. Alike the funnels, our droplet line element has as a boundary metric two copies of a Schwarzschild AdS$_4$ black hole.

Following \cite{deHaro:2000xn}, the $\chi-$dependent expectation value for the holographic stress energy tensor induced by the above coordinate transformation reads:
\begin{align}
& T_{t\phantom{t}}^{\phantom{t}t}=\frac{\ell^3_5}{16 \pi G_5  \rho ^2 \ell^4_4} \Bigg\{\left(1-\chi ^2\right)^4 \hat{\delta} (\chi )+(1+\rho ) \left(1-\chi ^2\right)^5 \left[2-3 (1+\rho ) \left(1-\chi ^2\right)\right]\nonumber
\\
&\qquad\qquad\qquad\qquad\qquad\qquad\qquad\qquad\qquad\qquad\qquad\qquad\qquad+\left(1-\chi ^2\right)^3(1+\rho ) \rho -\frac{3 \rho ^2}{4}\Bigg\}
   \\
& T_{\chi\phantom{\chi}}^{\phantom{\chi}\chi}=\frac{\ell^3_5}{16 G_5 \pi  \rho ^2 \ell_4^4} \Bigg\{\left(1-\chi ^2\right)^4 \hat{\epsilon} (\chi )+(1+\rho ) \left(1-\chi ^2\right)^5 \left[1-2 (1+\rho ) \left(1-\chi ^2\right)\right]-\frac{3 \rho ^2}{4}\Bigg\}
\\
& T_{\Omega_i\phantom{\Omega_j}}^{\phantom{\Omega_i}\Omega_j} =-\frac{1}{2}\left( T_{t\phantom{t}}^{\phantom{t}t}+ T_{\chi\phantom{\chi}}^{\phantom{\chi}\chi}+\frac{3 \ell _5^3}{16 \pi  G_5 \ell _4^4}\right)\delta_{\Omega_i\phantom{\Omega_j}}^{\phantom{\Omega_i}\Omega_j}
\end{align}
where $\Omega_i$ denotes any coordinate on the two sphere, $\hat{\delta}(\chi)=\delta(1-\chi)$ and $\hat{\epsilon}(\chi)=\epsilon(\chi)$.
\section{Diagnostics}
\label{sec:diag}
We will shortly compare the free energies of various droplets and funnels. However, since the bulk horizon extends to the boundary at infinity it clear that each of these quantities will be infinite.  As such, we need to find a consistent way of regularizing them to make the comparison meaningful. In a nutshell, we will use the uniform funnel as a regulator.

Let us first explain the regularization of the energy. The energy is defined in the usual way via the boundary stress energy tensor $T_{\mu\nu}$ \cite{Henningson:1998gx,Balasubramanian:1999re} (see \cite{Fischetti:2012rd,Marolf:2013ioa} for reviews aimed at relativists):
\begin{equation}
E \equiv -2\int_{\Sigma_t}\mathrm{d}^3\,x\,\sqrt{\sigma} T_{\mu\nu} K^{\mu} T^{\nu}\,,
\label{eq:energy}
\end{equation}
where $K$ is the static Killing vector $\partial_t$, $\Sigma_t$ is a surface of constant $t$ in the AlAdS boundary, with $\sigma$ and $T$ its induced metric and unit normal, respectively. The factor of two accounts for the fact that we have two copies of Schwarzschild-AdS$_4$ at the boundary. We can readily integrate in the angular coordinates, leaving only the integral over $X$.

For concreteness let us evaluate (\ref{eq:energy}) on the $d=4$ uniform funnel using eqs.~(\ref{eq:holographicstress}) with $\delta(\chi)=\epsilon(\chi)=0$.  We find
\begin{equation}
E_{\mathrm{uni}} =\frac{3 \ell_5^3\rho ^{3/2}}{4 G_5\ell_4} \lim_{\chi_1\to1^{-}}\int_{0}^{\chi_1}\mathrm{d}\chi\frac{\chi}{(1-\chi^2)^4}\, .
\end{equation}
As foreshadowed in section \ref{pert}, this is clearly divergent at the `equator' $\chi=1$ where the boundary metric fails to be $C^3$.  This occurs even though we use the standard `renormalized' boundary stress tensor $T_{\mu \nu}$  In order to deal with finite quantities, we will therefore systematically subtract the boundary stress-energy $T_{\mu\nu}^{\mathrm{uni}})$ of the uniform black funnel from both the non-uniform funnels and black droplets and compute
\begin{equation}
\Delta E \equiv -2\int_{\Sigma_t}\mathrm{d}^3\,x\,\sqrt{\sigma}(T_{\mu\nu}- T_{\mu\nu}^{\mathrm{uni}})K^{\mu} T^{\nu}\,.
\label{eq:regulatedenergy}
\end{equation}
We have explicitly checked that Eq.~(\ref{eq:regulatedenergy}) gives finite results for both the droplet and funnels phase. However, we stress that this is a very difficult quantity to compute because we must (numerically) cancel this apparent divergent behaviour. For this reason we had to resort to extended precision in our calculations. To ease our numerical scheme, we used a standard Newton-Raphson algorithm to find a solution with double precision, and switch to a Broyden type method when performing extended precision calculations.

The astute reader will notice that in order for $\Delta E$ to be finite for the droplet phase, we must have
\begin{equation}
\left.\delta(x)\right|_{x=1}\simeq -\frac{1}{2} \frac{\rho  (1+\rho )}{x}\,.
\end{equation}
We have checked this behaviour explicitly in our numerics. One can understand the origin of this simple pole by introducing funnel like coordinates near $(x,y)\sim(0,1)$ and solving the corresponding equations off the conformal boundary. One might wonder whether this divergence can be removed by an appropriate choice of conformal frame, but one can show it is not possible. One might also wonder how this behaviour is consistent with reflection symmetry around $x=0$. Indeed, near this special point it is natural to consider a ``radial'' variable $\tilde{r}\equiv \sqrt{x^2(2-x)^2+(1-y^2)^2}$, in terms of which all functions admit a regular Taylor series at least up to order $\tilde{r}^2$. For any $y\neq 1$, the reflection symmetry is present and that is why it is explicit in the bulk but not on the boundary.

The computation of the entropy has very similar issues: since the horizons intersect the boundary, they are non-compact, and as such have infinite area. An additional complication is that we must find a gauge invariant way to subtract the divergent piece, again using the uniform funnel as a reference. The induced metric, for both funnels and droplets, on the intersection of the horizon with a partial Cauchy slice of constant $t$ reads
\begin{equation}
\mathrm{d}s^2_{\mathcal{H}_t}=\frac{\ell_5^2}{1-y^2}\left[Q(y)\frac{\mathrm{d}y^2}{1-y^2}+W(y)\mathrm{d}\Omega_2^2\right]\,.
\end{equation}
For the uniform funnel we have $W=\rho$ and $Q=1$. Clearly, the area diverges near $y=1$ for all the phases, as anticipated. So we first introduce a cut-off $y_{i}$, where $i=\{\mathrm{D},\mathrm{F},\mathrm{UF}\}$ labels droplets, funnels and uniform funnels, respectively. The question is then how to relate the cut offs when we look at the difference in the areas. We do this my demanding that the radius of the $S^2$ (a gauge invariant quantity) match as the cut off recedes to the boundary. For the black funnel phase, such a procedure demands
\begin{equation}
y_{\mathrm{UF}}=\sqrt{1-\frac{1-y_{\mathrm{F}}^2}{S_2(0,y_{\mathrm{F}})}}\,.
\end{equation}
For the funnel phase, one finds after some work that this procedure yields
\begin{equation}
\Delta S_{\mathrm{F}} = 2\frac{\pi  \ell _5^3 \rho}{G_5}  \int_0^{1}\mathrm{d}y\frac{\sqrt{B(0,y)} S_2(0,y)-1}{\left(1-y^2\right)^2}
\end{equation}
with the factor of $2$ accounting for the two copies. In deducing this expression one has to ensure that the limit $y_{\mathrm{F}}\to1^-$ exists and is finite, which one can do using the boundary expansions determined previously.

The droplet phase is more complicated, but the result is the same in spirit. To match the spheres we now have
\begin{equation}
y_{\mathrm{UF}}=\sqrt{1-\frac{(1-y_{\mathrm{D}}^2)\rho}{y_{\mathrm{D}}S_2(1,y_{\mathrm{D}})}}\,,
\end{equation}
and we wish to look at
\begin{align}
\Delta S_{\mathrm{D}} & = \frac{2\pi\ell_5^3}{G_5}\lim_{y_{\mathrm{D}}\to1^-} \left[\int_0^{y_D} \mathrm{d}y \frac{y^2 \sqrt{B(1,y)} S_2(1,y)}{\left(1-y^2\right)^2}-\int_0^{y_{\mathrm{UF}}}\mathrm{d}\tilde{y}\frac{ \rho}{\left(1-\tilde{y}^2\right)^2}\right]\nonumber
\\
& = \frac{2\pi\ell_5^3}{G_5}\lim_{y_{\mathrm{D}}\to1^-} \Bigg\{\int_0^{y_D}\mathrm{d}y\left[\frac{y^2 \sqrt{B(1,y)} S_2(1,y)}{\left(1-y^2\right)^2}-\frac{1}{\left(1-y^2\right)^2}\left[1-\frac{1}{2} (1-\rho ) \left(1-y^2\right)\right]\right]\nonumber
\\
&+\int_0^{y_D}\mathrm{d}y\frac{1}{\left(1-y^2\right)^2}\left[1-\frac{1}{2} (1-\rho ) \left(1-y^2\right)\right]-\int_0^{y_{\mathrm{UF}}}\mathrm{d}\tilde{y}\frac{ \rho}{\left(1-\tilde{y}^2\right)^2}\Bigg\}\nonumber
\\
& = \frac{2\pi\ell_5^3}{G_5}\Bigg\{\int_0^{1}\mathrm{d}y\left[\frac{y^2 \sqrt{B(1,y)} S_2(1,y)}{\left(1-y^2\right)^2}-\frac{1}{\left(1-y^2\right)^2}\left[1-\frac{1}{2} (1-\rho ) \left(1-y^2\right)\right]\right]+\frac{\rho  \log \rho }{4}\Bigg\}\,,
\end{align}
where we have explicitly checked (by using our asymptotic expansion (\ref{eqs:expansiony1}) around $y=1$) that the remaining integral is finite and can be readily evaluated using our numerical data.

To compute the different in Helmoltz free energies we simply take $\Delta F = \Delta E-T\Delta S$. This is the appropriate quantity to study when exploring the phase diagram at constant temperature, as we shall do below.

Finally, we will be using isometric embeddings when comparing funnels and droplets. These are specially useful if we want to visualize where the horizon bulges out. In a nutshell, we will embed the spatial cross section of the horizon into four dimensional hyperbolic space. We will foliate four-dimensional hyperbolic space using three-dimensional hyperbolic space:
\begin{equation}
\mathrm{d}s^2_{\mathbb{H}} = \frac{\tilde{\ell}^2_4}{1-\tilde{Y}^2}\left\{\frac{\mathrm{d}\tilde{Y}^2}{1-\tilde{Y}^2}+\frac{1}{\tilde{\ell}_3^2}\left[\frac{\mathrm{d}\tilde{R}^2}{1+\frac{\tilde{R}^2}{\tilde{\ell}_3^2}}+\tilde{R}^2 \mathrm{d}\Omega^2_2\right]\right\}\,,
\end{equation}
where $\tilde{\ell}_3$ and $\tilde{\ell}_4$ are the hyperbolic length scales of the embedding space. One then searches for an embedding of the form $(\tilde{R}(y),Y(y))$, which gives the following induced metric
\begin{equation}
\mathrm{d}\hat{s}^2_{\mathbb{H}} = \frac{\tilde{\ell}^2_4}{1-Y(y)^2}\left\{\left[\frac{\tilde{Y}^\prime(y)^2}{1-\tilde{Y}(y)^2}+\frac{1}{\tilde{\ell}_3^2}\frac{\tilde{R}^\prime(y)^2}{1+\frac{\tilde{R}(y)^2}{\tilde{\ell}_3^2}}\right]\mathrm{d}y^2+\frac{\tilde{R}(y)^2}{\tilde{\ell}_3^2}\mathrm{d}\Omega^2_2\right\}\,.
\end{equation}
We can now compare this line element with the metric of a funnel or droplet induced on the intersection of the horizon with a partial Cauchy surface of constant $t$ and read off a nonlinear first order equation for $\tilde{Y}(y)$. We fix the boundary conditions by demanding $\tilde{Y}(0)=0$ and choose the ratio
\begin{equation}
\frac{\tilde{\ell}_3}{\tilde{\ell}_4}\frac{\ell_5}{\ell_4}
\end{equation}
so that $\tilde{R}(1)=\sqrt{\rho}\,\ell_4$. The curve traced by $(\tilde{R}(y),\tilde{Y}(y))$, as we vary $y\in(0,1)$, is the embedding diagram. Note that in this embedding, a uniform funnel is simply given by $\tilde{R}(y)=\sqrt{\rho}\,\ell_4$, as expected.
\section{Droplets and funnels for $d = 4$}
\label{sec:results}

We now present the results of our numerics. We also encourage the reader to consult appendix \ref{ap:2} for evidence that our simulations exhibit the expected convergence with increasing numbers of grid points.   We begin with the black funnels and then include droplets below.  As described in section \ref{pert calcs}, the uniform black funnel becomes unstable for $\rho > \rho_{\mathrm{onset}}$.  In this regime, at least perturbatively close to $\rho_{\mathrm{onset}}$, we can identify additional fat and thin branches of black funnels in our numerics.  The former branch exists at $\rho < \rho_{\mathrm{onset}}$, while the latter exists at $\rho > \rho_{\mathrm{onset}}$.  In fact, as is clear from the perturabtive discussion, the fat and thin funnels are really the same branch of (nonuniform) funnel solutions continued to opposite sides of $\rho_{\mathrm{onset}}$.  We can then follow these branches numerically to larger values of $|\rho - \rho_{\mathrm{onset}}|$.

\begin{figure}
\centerline{
\includegraphics[width=\textwidth]{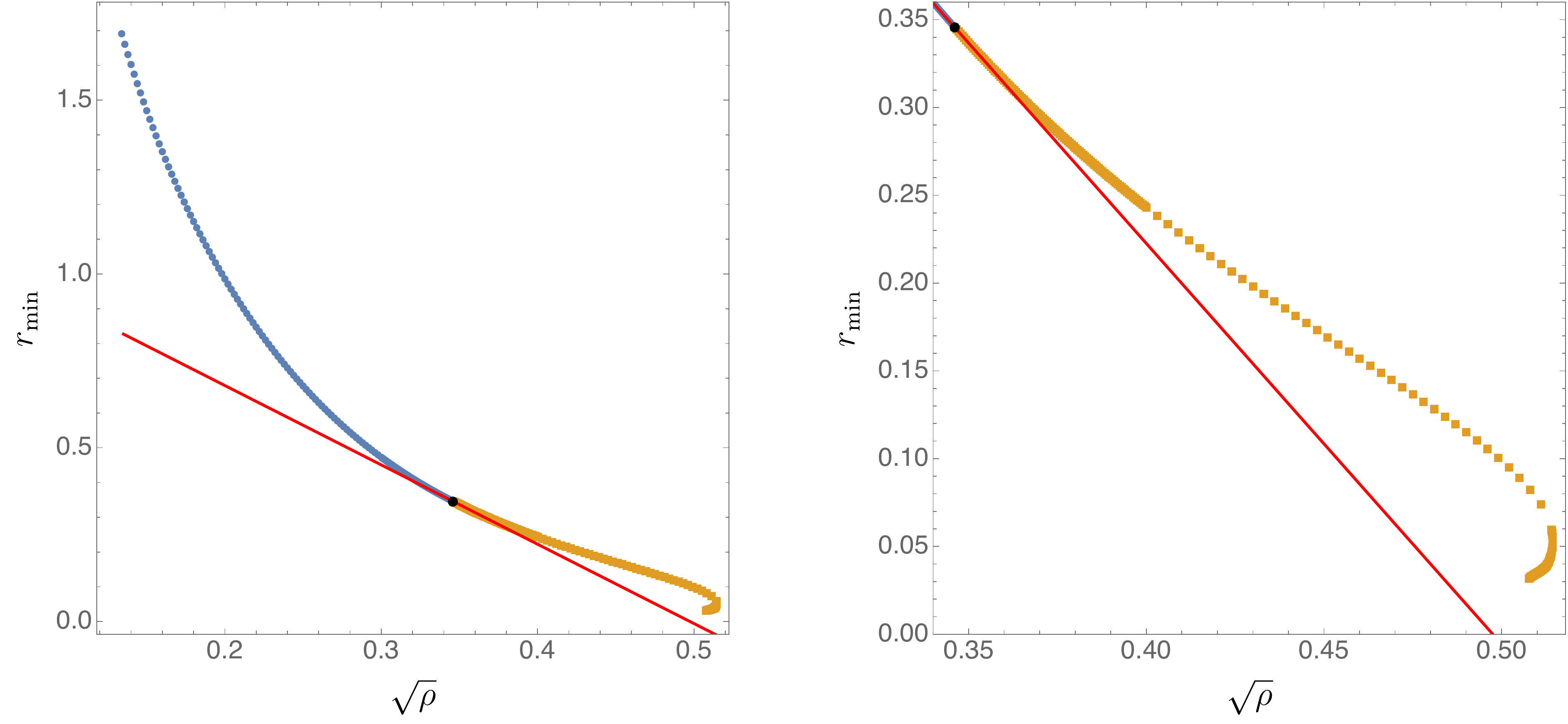}
}
\caption{Minimum size of the $S^2$ along the funnel horizon for both fat (blue disks) and thin (orange squares) funnels. The black disk indicates the onset of the Gregory-Laflamme-like instability afflicting the uniform phase, corresponding to $\rho=\rho_{\mathrm{onset}}$. The right panel shows an expanded view of the region near $\rho_{\mathrm{turn}}\approx 0.264553$.  The line predicted by our 2nd order perturbative calculation is also shown (solid red line).}
\label{fig:bulge}
\end{figure}

As a measure of how fat or thin the funnels may be, we can monitor the minimum radius $r_{\min}\equiv \sqrt{\rho\,S_2(0,0)}$ of the $S^2$ along the horizon as a function of $\sqrt{\rho}$.  As shown in figure \ref{fig:bulge},  the thin branch reaches a turning point
at $\rho_{\mathrm{turn}}\approx 0.264553$ (where $r_{\min}\approx 0.0534059$) at which $\rho$ then begins to decrease with continued decrease of $r_{\min}$, and beyond which funnels appear not to exist.
This non-monotonic behaviour is very reminiscent of the behaviour of the transition between black strings and localised black holes in five and six dimensions, recently observed in \cite{Kalisch:2015via,Kalisch:2016fkm,Kalisch:2017bin}. We shall see below that the droplet phase also shows hints of merging with the thin funnel phase for values of $\rho$ in the vicinity of $\rho_{\mathrm{turn}}$ in direct parallel with known results \cite{Kol:2002xz,Sorkin:2003ka,Kudoh:2003ki,Kudoh:2004hs,Kol:2005vy,Headrick:2009pv,Kalisch:2015via,Kalisch:2016fkm,Kalisch:2017bin}for Kaluza-Klein black holes.

\begin{figure}
\centerline{
\includegraphics[width=0.6\textwidth]{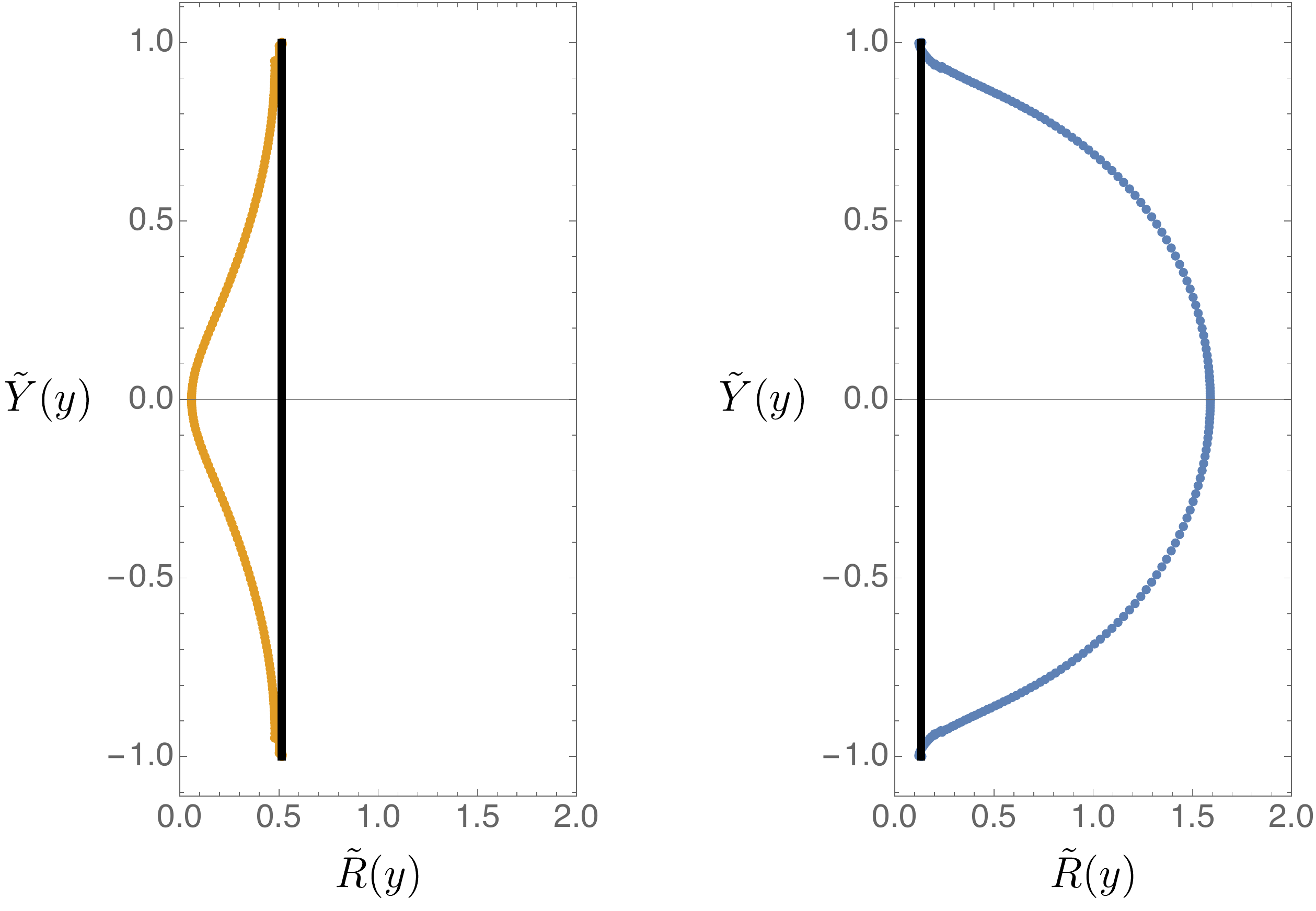}
}
\caption{Embedding diagram in the hyperbolic plane for a sample thin funnel with $\sqrt{\rho}=0.134$ (left panel), and a sample fat funnel with $\sqrt{\rho}=0.5143$ (right panel). The black solid vertical line in each panel indicates the embedding diagrams of the uniform funnels with the same value of $\rho$.}
\label{fig:embefunnels}
\end{figure}

We can also see the predicted outward or inward bulging of the horizon by looking at the isometric embeddings found in our previous section. These are plotted in Fig.~\ref{fig:embefunnels} for the smallest and largest values of $\rho$ that we managed to probe.

\begin{figure}
\centerline{
\includegraphics[width=0.4\textwidth]{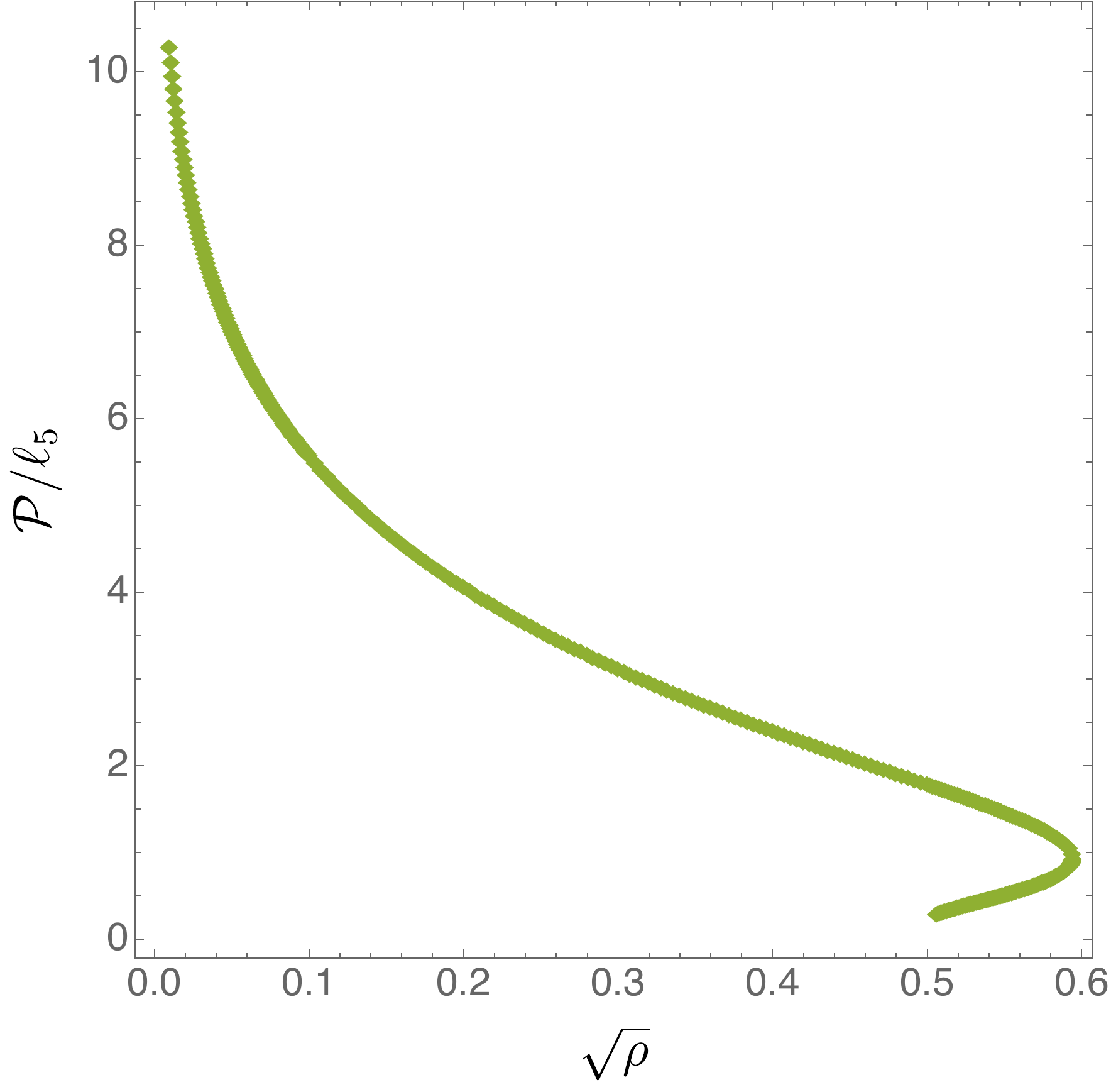}
}
\caption{The proper distance $\mathcal{P}$ between the two horizons in the droplet phase as a function of $\sqrt{\rho}$.}
\label{fig:properdistance}
\end{figure}

For the droplets, we decided to monitor the proper distance $\mathcal{P}$ along the $S^2$ axis between the two horizons. The results are shown in \eqref{fig:properdistance}.  Our $\mathcal{P}$ diverges as $\rho \rightarrow 0$ since the horizons recede to the boundary in that limit.  The quantity $\mathcal{P}$ then generally decreases with $\rho$, though we again find a turning point $\sqrt{\rho_\mathrm{turn,D}} \approx0.5943$ beyond which continued decrease of $\mathcal{P}$ requires $\rho$ to decrease.  Droplets appear not to exist for $\rho > \rho_\mathrm{turn,D}$.  When two droplets exist, we may call them short and long in analogy with the $d=3$ case \cite{Hubeny:2009rc} reviewed in section \ref{BTZ}.  As in that case, the short droplets have smaller $\Delta F$ as may be seen by comparing figures \ref{fig:properdistance} and \ref{fig:freeenergy}. The behavior is again akin to that of localized black holes in Kaluza-Klein theory
\cite{Wiseman:2002ti,Kudoh:2003ki,Kudoh:2004hs,Headrick:2009pv,Dias:2017uyv,Kalisch:2017bin,Kalisch:2018efd,Cardona:2018shd,Ammon:2018sin}. 

\begin{figure}
\centerline{
\includegraphics[width=\textwidth]{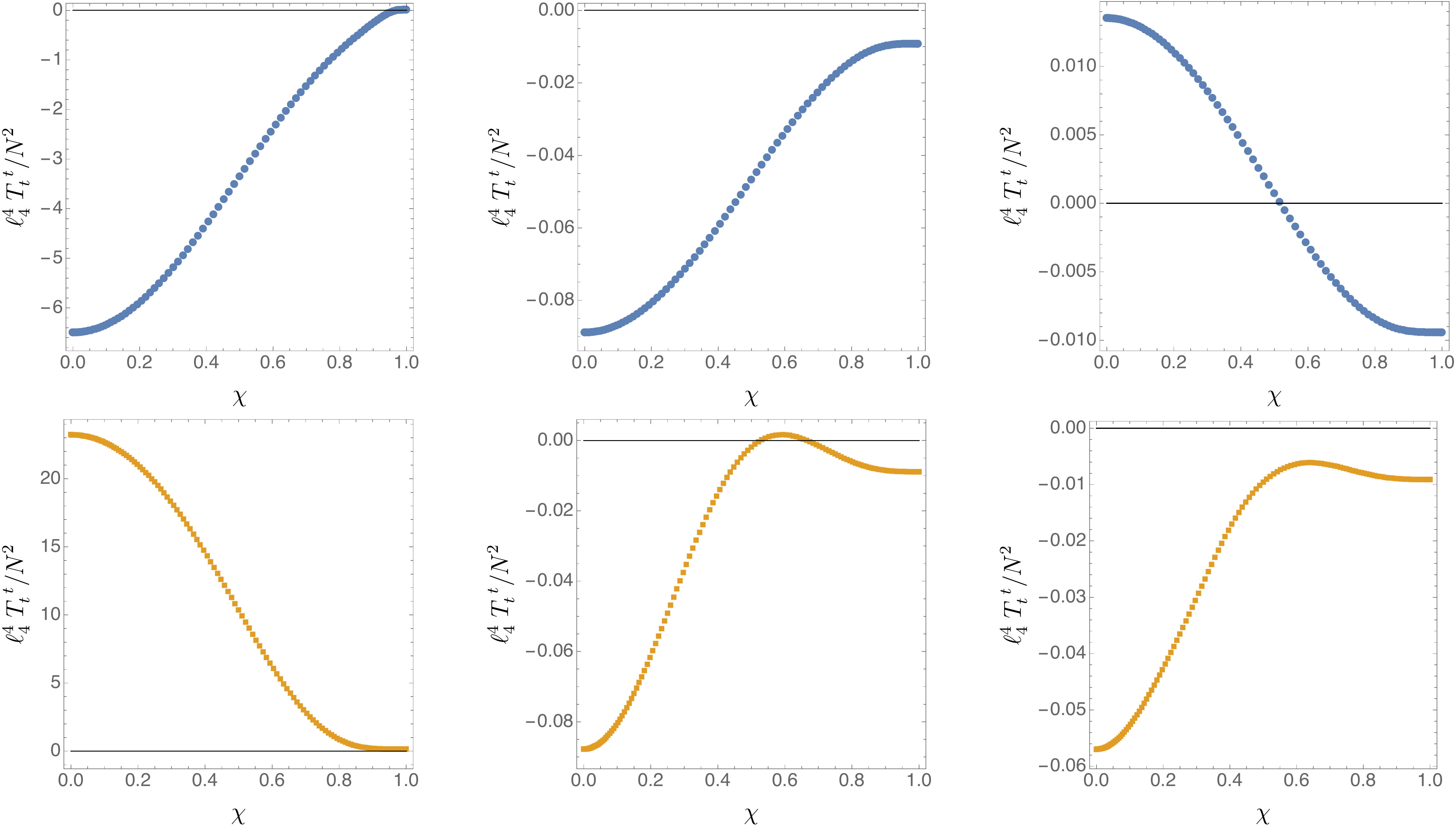}
}
\caption{Boundary stress energy tensors for funnels (top row) and droplets (bottom row). On the top row we show results for two fat funnels with $\sqrt{\rho}=0.134, 0.3$ and an upper branch (in the sense of figure \ref{fig:bulge}; \emph{i.e.}, less thin) thin funnel with $\sqrt{\rho}=0.5143$, from left to right, while for the bottom row we show the (unique)  droplets with $\sqrt{\rho}=0.01, 0.3$ as well as our longer (smaller ${\cal P}$) droplet with $\sqrt{\rho}=0.5207$.}
\label{fig:tensors}
\end{figure}

Other properties of interest include profiles of the boundary stress-energy tensor along the AlAdS boundary in each phase.  Due to spherical symmetry and time-reversal symmetry, this tensor has only three non-zero components.  Furthermore, the trace anomaly fixes its trace in terms of the boundary curvature, and covariant conservation imposes a second constraint.  As a result, there is only one independent component, with any two being determined by the third.  We plot $T_{t}^{\phantom{t}t}$ in Fig.~\ref{fig:tensors} as a function of the boundary coordinate $\chi$ for representative cases of interest again using the parameter $N^2$ from \eqref{eq:N2}. For small enough boundary black holes, the fat funnel phase has a very negative $T_{t}^{\phantom{t}t}$, while the droplet phase becomes increasingly positive away from $\chi=1$.  At the largest value of $\rho$ shown the quantity  $T_{t}^{\phantom{t}t}$ becomes negative everywhere for the droplet, while in other cases the sign depends on $\chi$. For all shown phases we find $\lim_{\chi\to1^-}\ell_4^4T_{t}^{\phantom{t}t}/N^2=-3/(32\pi^2)<0$ as required for finitenes of $\Delta E$ in Eq.~(\ref{eq:regulatedenergy}).

However, the most important quantities for each solution are the total entropy, energy and free energy, \emph{i.e.} thermodynamic properties.  As described in section \ref{2nd}, the dominant phase for each boundary metric is the one minimizing the free energy, and thus $\Delta F$.   For completeness, we nevertheless first show $\Delta E$ and $\Delta S$ separately in figure \ref{fig:entropy_full}, \ref{fig:comparison_fat_schw}, and \ref{fig:energy} below.
These quantities satisfy a first law in the usual sense:
\begin{equation}
\mathrm{d}\Delta E = T\mathrm{d}\Delta S\,,
\label{eq:firstlaw}
\end{equation}
which follows directly from \cite{Wald:1993nt,Iyer:1994ys,Iyer:1995kg,Wald:1999wa}. We have checked that our numerical data satisfies this very stringent relation, and we use it as a numerical check. All solutions presented in this manuscript satisfy this form of the first law, with a relative error of less than $1\%$. In all thermodynamics plots we  represent droplets by orange squares and funnels by blue disks.

We also comment that the geometry of a fat funnel near its central bulge
remarkably like that a round spherical black hole.  To make this quantitative, figure \ref{fig:comparison_fat_schw} compares the the entropy of a Schwarzschild-AdS$_5$ black hole with $\Delta S$ for a very fat funnel at the same temperature.  Recalling that fat funnels exist only for small $\rho$, one sees that at high temperature subtracting the uniform funnel entropy in $\Delta S$ will have little effect on the finite part of the entropy so that this comparison will be meaningful up to some constant offset of order 1. Note that the plot for fat funnels is completely consistent with the SAdS$_5$ result $\Delta S\propto T^3$ at large $T$.

\begin{figure}
\centerline{
\includegraphics[width=\textwidth]{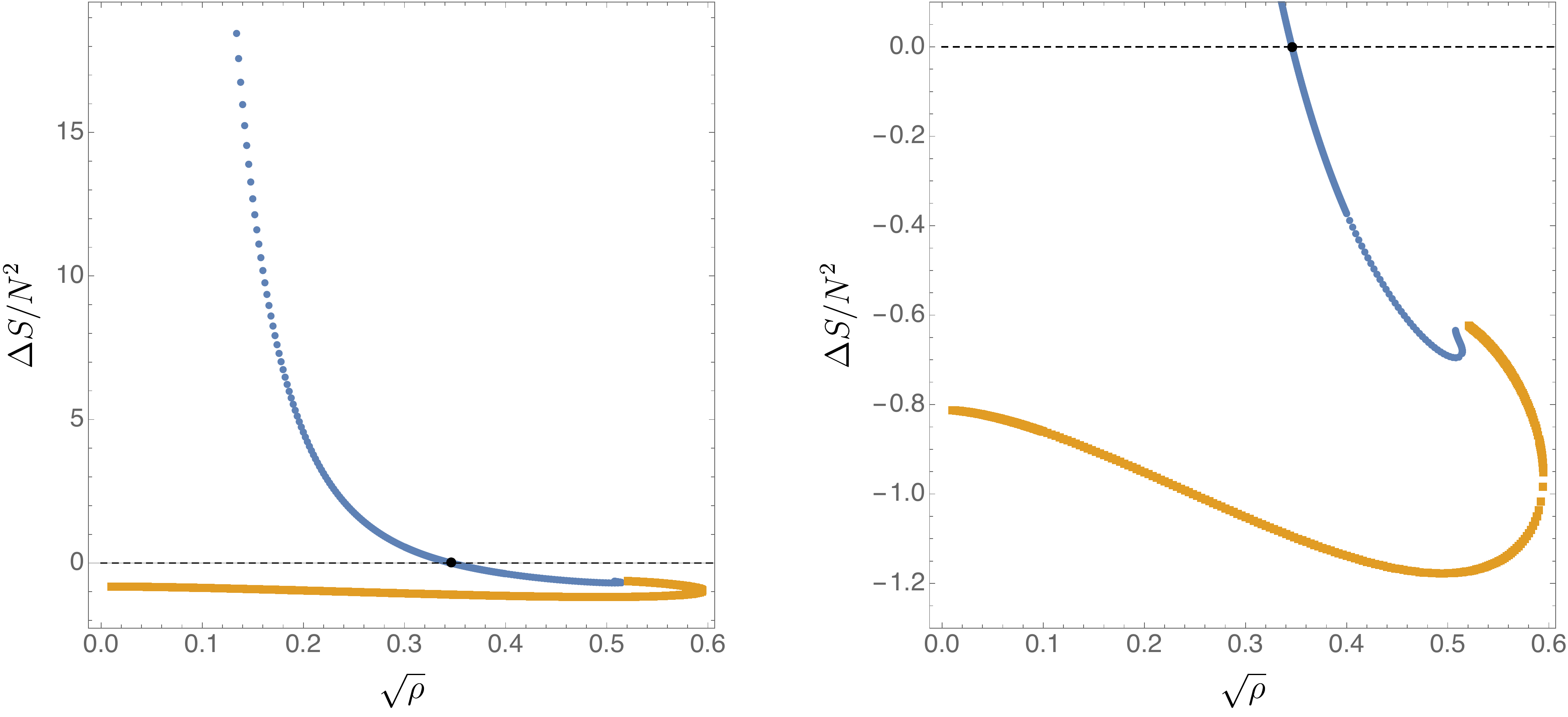}
}
\caption{The entropy of various phases with fixed SAdS$_4$ boundary metric parametrised by $\rho$: the orange squares represent the droplets and the blue disks the funnels. The right panel shows a close-up of the region close to the merger point where the funnel and droplet families appear to meet. The black disk represents the transition between stable and unstable non-uniform funnels.}
\label{fig:entropy_full}
\end{figure}

\begin{figure}
\centerline{
\includegraphics[width=0.4\textwidth]{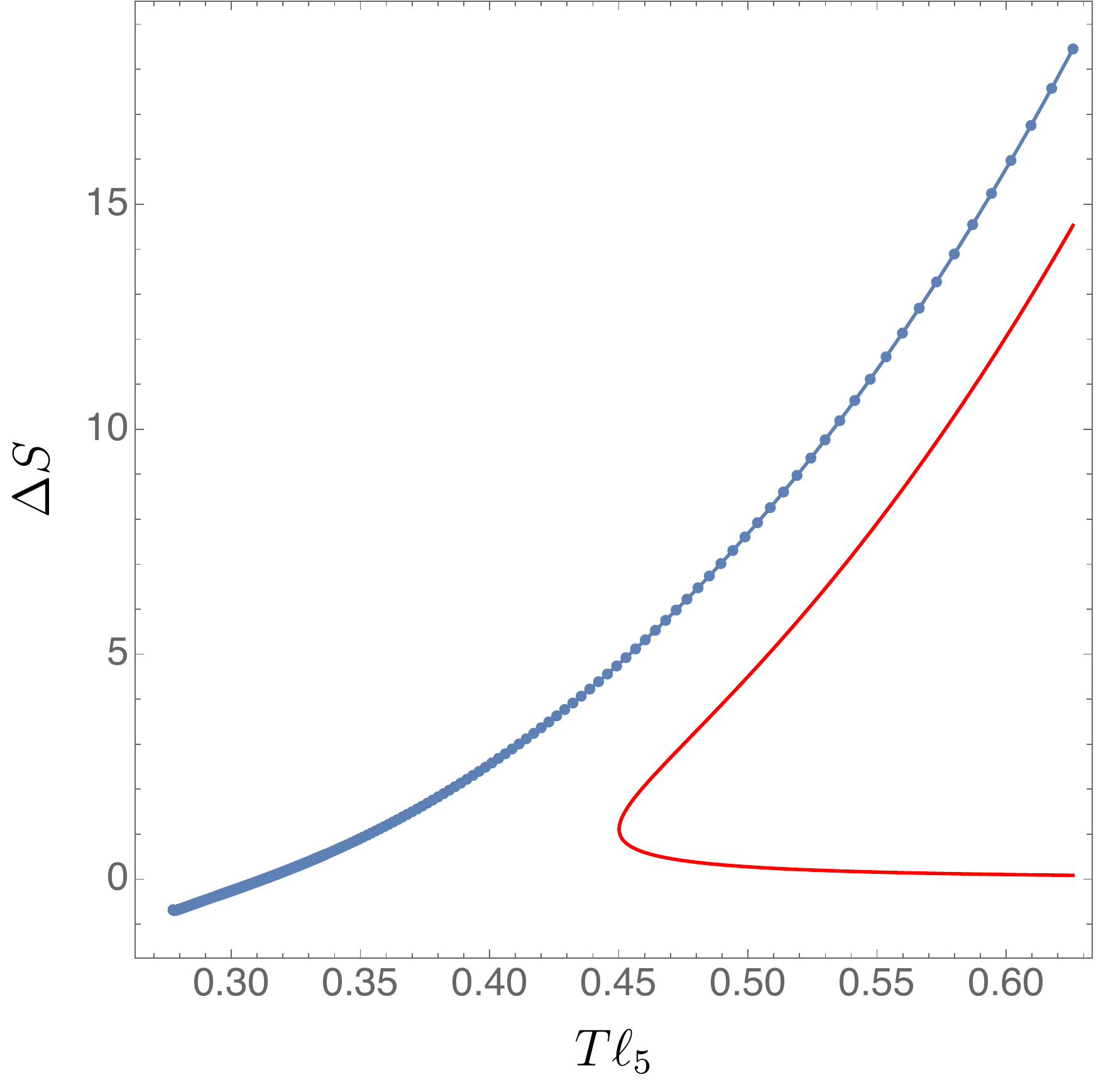}
}
\caption{Comparing $\Delta S$ for a fat funnel with the entropy of a spherical Schwarzschild-AdS$_5$ black hole at the same temperature: the solid red line corresponds to the Schwarzschild-AdS$_5$  black hole, and the blue disks to the fat funnel phase. Both phases exhibit a power law behaviour at large values of $T\ell_5$.}
\label{fig:comparison_fat_schw}
\end{figure}

\begin{figure}
\centerline{
\includegraphics[width=\textwidth]{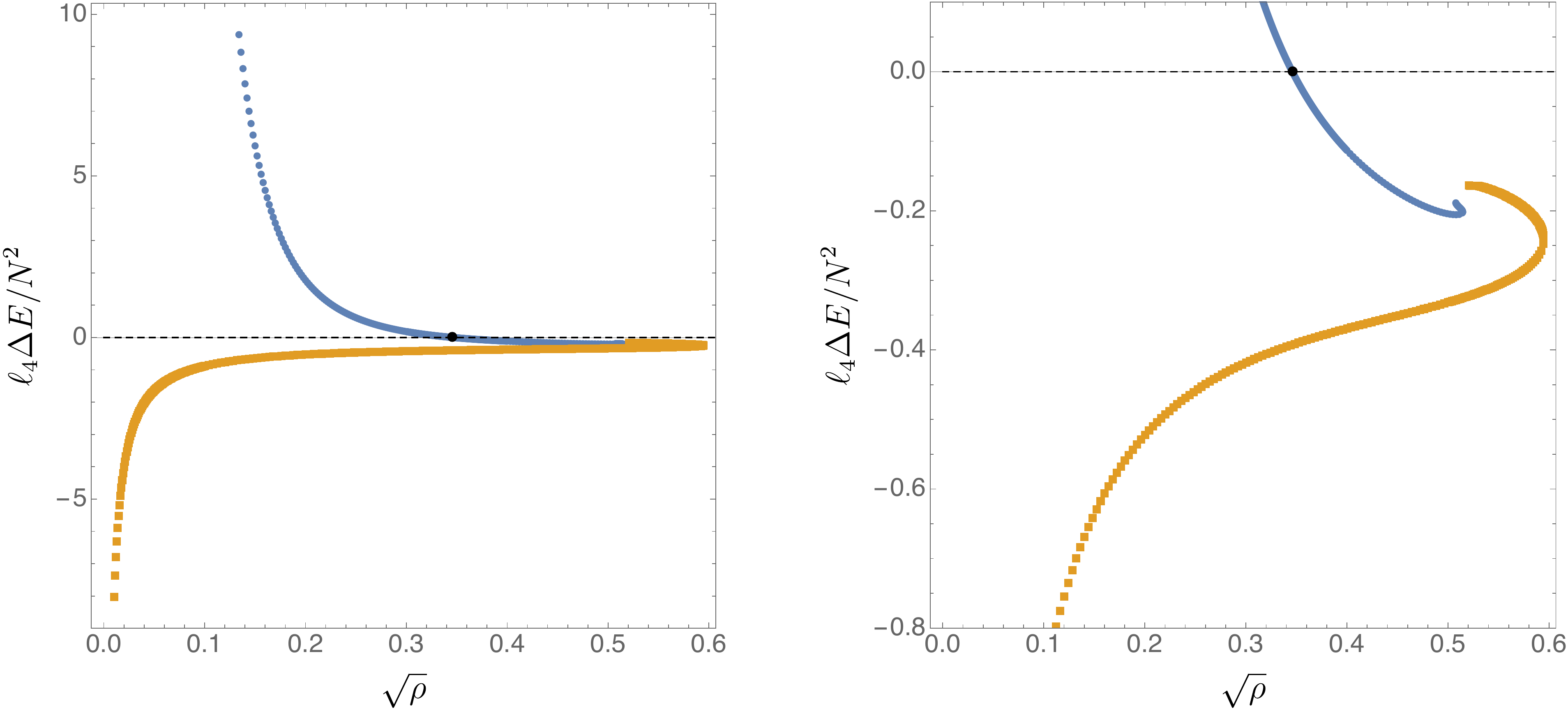}
}
\caption{The energy of various phases with fixed SAdS$_4$ boundary metric parametrised by $\rho$: the orange squares represent the droplets and the blue disks the funnels. The right panel shows a close-up of the region close to where funnels and droplets appear to merge. The black disk represents the transition between stable and unstable non-uniform funnels.}
\label{fig:energy}
\end{figure}

\begin{figure}
\centerline{
\includegraphics[width=0.4\textwidth]{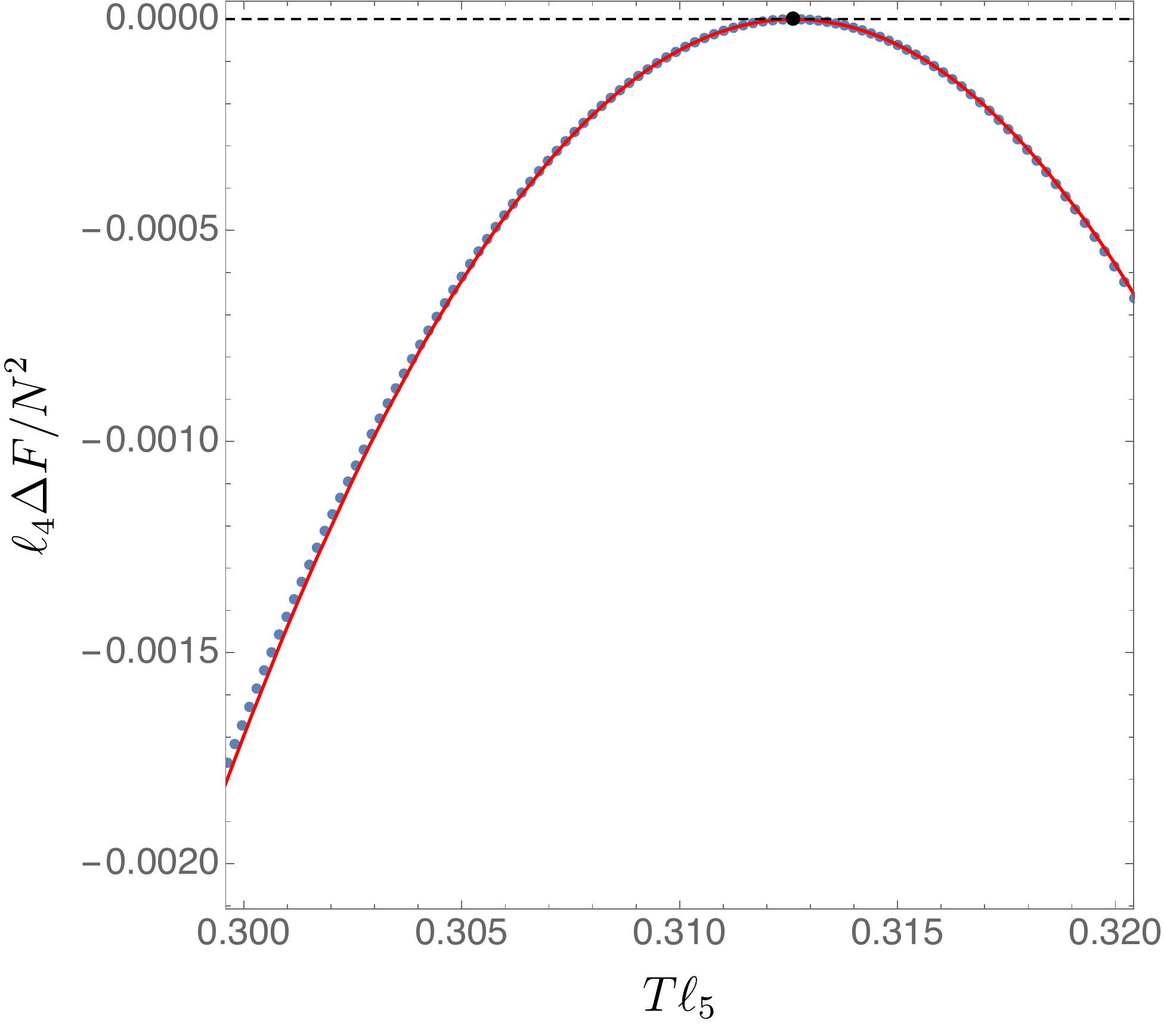}
}
\caption{Free energy of the fat and thin funnels as a function of $T\ell_5$ computed near the merger point: the red solid line corresponds to the nonlinear perturbative result (\ref{eq:pert}) and the blue disks to our exact numerical data. The agreement between the two methods is reassuring.}
\label{fig:compa}
\end{figure}

\begin{figure}
\centerline{
\includegraphics[width=\textwidth]{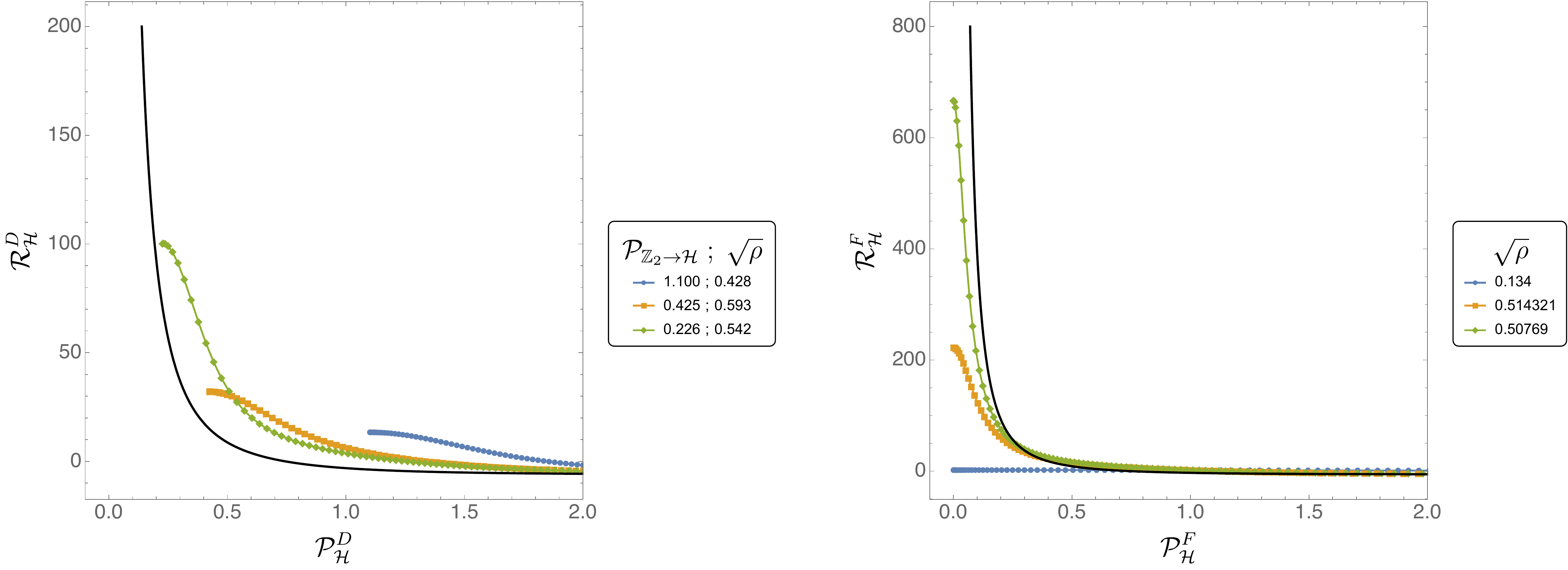}
}
\caption{Comparisons of the horizon geometries in our numerical solutions with that of \eqref{AdSDC}.  The quantity $\mathcal{P}_{\mathcal{H}}^F$ refers to proper distance along the funnel horizon measured from the plane of $\mathbb{Z}_2$ symmetry, while $\mathcal{P}_{\mathcal{H}}^D$ measures proper distance along a curve that first runs from the plane of $\mathbb{Z}_2$ symmetry to the droplet and then along the droplet horizon.  The quantities  $\mathcal{R}_{\mathcal{H}}^F$,  $\mathcal{R}_{\mathcal{H}}^D$ denote Ricci scalars at corresponding horizon points.  We compare $\mathcal{P}_{\mathcal{H}}^F$ vs. $\mathcal{R}_{\mathcal{H}}^F$ and $\mathcal{R}_{\mathcal{H}}^D$ vs $\mathcal{R}_{\mathcal{H}}^D$ with corresponding quantities (solid lines) computing on the horizon of \eqref{AdSDC}.  The annotations show the relevant values of $\sqrt{\rho}$ and, for the droplets, the proper distance along the rotation axis from the symmetry plane and the droplet tip (which is a natural measure of the distance of any given solution from the transition).
}
\label{fig:DC}
\end{figure}

\begin{figure}
\centerline{
\includegraphics[width=\textwidth]{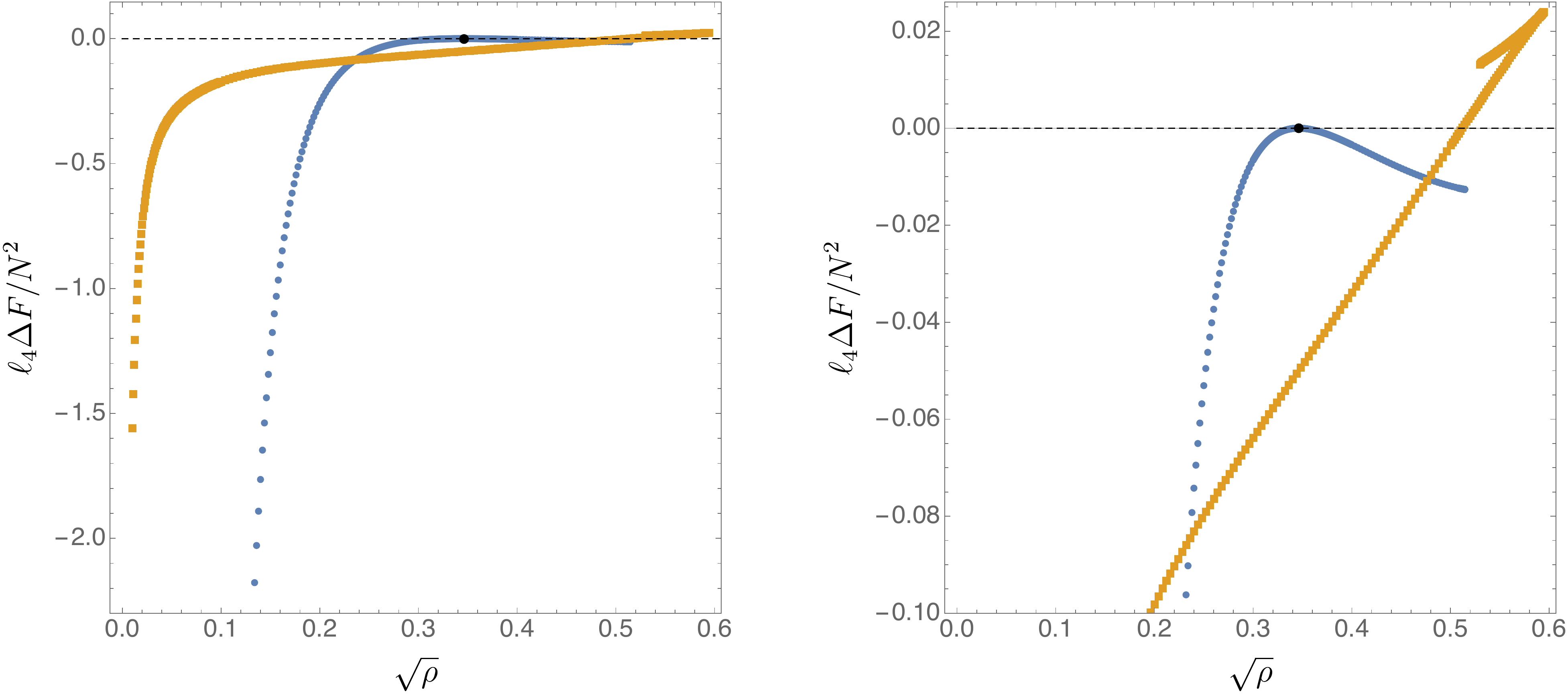}
}
\caption{Free energy of all all the competing phases with fixed boundary metric parametrised by $\rho$: the orange squares represent the droplets and the blue disks the funnels. The right panel shows a close-up of the region close to where non-uniform and uniform funnels merge.  The finer scale used in the close up makes the merger less apparent, but given the results of figure \ref{fig:DC} we expect that further numerics pushing closer to the presumed merger would fill in the gap between the droplet and funnel free energies.  The black disk represents the transition between stable and unstable non-uniform funnels.  The close-up also includes the region where droplets are expected to merge with funnels, and in particular the turning point $\rho_{\mathrm{turn}}$ for the thin funnels.  This turning point was easily visible in figures \ref{fig:entropy_full} and \ref{fig:energy}, showing finite differences $S_{\mathrm{thin}}-S_{\mathrm{thinner}}<0$ and $E_{\mathrm{thin}}-E_{\mathrm{thinner}}<0$ when comparing two branches of thin funnels at the same $\rho$. But the contributions nearly cancel when comparing free energies, so that the two parts of the thin funnel curve here are separated by much less than the width of the blue disks and the presence of two solutions near $\rho_{\mathrm{turn}}$ cannot be seen; i.e., as the funnels become even thinner they are now moving toward smaller $\rho$ and nearly retracing the curve defined by their somewhat-less-thin relatives.  However, there is a tiny difference between the free energies of the two thin funnels at any given value of $\rho$.  In particular, numerical computations find $F_{\mathrm{thin}}-F_{\mathrm{thinner}} <0$ as expected.
}
\label{fig:freeenergy}
\end{figure}

As noted in the figure captions, the limit of very long droplets appears to merger with the limit of very thin funnels.  In particular, the free energies of these families of solution appear to agree in that limit.  Such a merger would again be in direct parallel with our understanding of Kaluza-Klein black holes.  Indeed, as argued by Kol \cite{Kol:2002xz,Kol:2005vy}, the limiting solution can be expected to take the local form of a double cone.  In the AdS$_5$ context one can find an exact analytic (Lorentz-signature) double-cone metric
\begin{equation}
\label{AdSDC}
\mathrm{d}s2 = \mathrm{d}\rho^2+ H \mathrm{d}\Omega_2^2+ H \mathrm{d}\tilde{\Omega}_2^2
\end{equation}
with $H = \frac{L^2}{3} \sinh^2\left(\frac{\rho}{L}\right)$, $\mathrm{d}\Omega_2^2$ the line element on the unit metric $S^2$, and $\mathrm{d}\tilde{\Omega}_2^2$ the line element on a unit-radius 1+1 de Sitter space.  While the boundary metric of this AlAdS spacetime does not match any of the ones we wish to consider, that does not prevent our solutions from approaching \eqref{AdSDC} {\it locally} in the bulk near the point where the horizons merge or pinch off.  Comparisons of the horizon geometries in our numerical solutions with that of \eqref{AdSDC} are shown in figure \ref{fig:DC}.  These plots are consistent with the idea that the funnels and droplets do indeed merge at a double-cone-like solution.

Finally, we discuss the free energies in detail. Our boundary conditions are specified by a single parameter $\rho$, which in turn determines the temperature $T(\rho)$ of the boundary black holes.  Since our bulk solutions are in equilibrium with the boundary black holes, $T(\rho)$ is also the temperature of our bulk horizons.   As a check on our computations, figure \ref{fig:compa} compares numerical results for $\Delta F$ for fat and thin funnels near the transition with the 2nd order perturbative result \eqref{eq:pert} and finds strong agreement.

The free energies plotted in figure \ref{fig:freeenergy} indicate a rather intricate series of first order phase transitions as follows:  For $\sqrt{\rho} \lesssim 0.236037$ the most dominant phase we study is a fat funnel.  This is as expected from figure \ref{fig:BHlines} (right) since droplet-plus-black-hole phases are beyond the scope of this work.    At intermediate $\rho$ ($0.236037<\sqrt{\rho}<0.477205$), the most dominant phase is a droplet.  Here figure \ref{fig:BHlines} makes no definite prediction.  At somewhat larger $\rho$ ($0.477205<\sqrt{\rho}<\sqrt{\rho_{\mathrm{turn}}} \approx 0.5143$) there is a small region where our most dominant phase is a thin funnel.  But as discussed above there appear to be no thin funnels for $\rho > {\rho_{\mathrm{turn}}}$.  Instead, in this final regime the dominant phase we study is the uniform funnel.  While figure \ref{fig:BHlines} makes no definite prediction for the dominant phase beyond the small $\rho$ regime, all of these results are certainly compatible with the general analysis of section \ref{DFgen}.

A notable feature, however, is that the most dominant phase in figure \ref{fig:freeenergy}
for $\rho \lesssim \rho_\mathrm{turn}$ (the thin funnel) does not exist beyond $\rho_\mathrm{turn}$.  If there were no additional phases beyond those shown in the figure, increasing $\rho$ slightly beyond $\rho_{turn}$ would thus cause $\Delta F$ to increase by a finite amount.
As described in section \ref{sec:disc}, this would violate the 2nd law.  As a result, there must be additional phases in this regime.  See further discussion in section \ref{sec:disc}

\section{Discussion}
\label{sec:disc}

In the above work, we constructed and analyzed pure funnel and droplet AlAdS solutions numerically with $d=4$ Schwarzschild-AdS boundaries.  For each such boundary metric, the solution with lowest free energy $F= E-TS$ should be considered most dominant, as the second law for such solutions \cite{Bunting:2015sfa} will prevent it from decaying to any solution with higher free energy.   Here the analytic SAdS black string solution of e.g. \cite{Gregory:2008br} is interpreted as a funnel phase, which we call the uniform funnel due to its conformal symmetry.  Nonuniform funnels branch off from this uniform funnel at the point where the uniform funnel becomes unstable.  This process is closely akin to the Gregory-Laflamme instability of black strings \cite{Gregory:1993vy} and our phase diagram \ref{fig:freeenergy} strongly resembles that of Kaluza-Klein black holes \cite{Wiseman:2002ti,Kudoh:2003ki,Kudoh:2004hs,Headrick:2009pv,Dias:2017uyv,Kalisch:2017bin,Kalisch:2018efd,Cardona:2018shd,Ammon:2018sin}.
Here the funnels play the role of black strings and black droplets play the role of Kaluza-Klein black holes.

Taking $R$ to be the radius of our SAdS boundary black holes, we find so-called `fat' non-uniform funnels to dominate at small $R$, droplets to dominate for intermediate values of $R$, and uniform funnels to dominate at large $R$; see figure \ref{fig:freeenergy}.  Horizon shapes, energies, entropies, stress-tensor profiles were explored in section \ref{sec:results} and -- with one exception described below -- fit with general expectations reviewed in sections \ref{DFgen} and \ref{pert}, and also with perturbative funnel computations performed in section \ref{pert calcs}.  This supports the intuition developed in section \ref{overview} and suggests it will continue to hold for other boundary metrics. In particular, it supports the view of fat funnels as essentially large global SAdS black holes attached to the boundary by small pieces of uniform funnels; see e.g. figure \ref{fig:comparison_fat_schw}.  One interesting surprise captured by both numerics and perturbation theory is that even for $d=4+1$ bulk dimensions, non-unform funnels have lower free energy than uniform funnels even close to the onset of the uniform funnel instability.  In contrast, in the perturbative regime non-uniform black strings with Kaluza-Klein boundary conditions dominate over the corresponding uniform (translationally-invariant) string only for bulk dimension $d+1 = 14$ or greater \cite{Sorkin:2004qq}.

The above-mentioned exception concerns thin funnels, which we find to dominate in a regime between those dominated by droplets and that dominated by uniform funnels.  In particular, they have lower free energy than our other phases near the point $\rho_{\mathrm{turn}} = R^2_{\mathrm{turn}}/\ell^2_4$ where the line of thin funnels turns around.  However, the second law of thermodynamics requires that there be an additional more-dominant phase in at least the region very close to $\rho_{\mathrm{turn}}$. This may be argued by supposing that we begin with a thin funnel at $\rho$ slightly less that $\rho_{\mathrm{turn}}$ and dynamically change the boundary metric to increase $\rho$.  The free energy $\Delta F$ should be a continuous function of time, so even at $\rho$ slightly more than $\rho_{\mathrm{turn}}$ our $\Delta F$ must remain lower than that of the uniform funnel.  But as discussed in section \ref{2nd}, the 2nd law \cite{Bunting:2015sfa} for our systems prohibits $\Delta F$ from increasing when the boundary metric is static.  Our system is thus forbidden from evolving to the uniform funnel\footnote{Strictly speaking the 2nd law of \cite{Bunting:2015sfa} assumes an appropriate version of cosmic censorship, which as we shortly discuss seems likely to fail.  As with the original arguments of Gregory and Laflamme \cite{Gregory:1993vy}, we thus assume an appropriate extension to a full theory of quantum gravity that can accommodate any singularities that arise.}.  On the other hand, the system is clearly dissipative, so it should settle down to a stationary solution of lower free energy.

In considering possible candidates for this new phase, one should note that the funnel and droplet phases we study appear to merge in this regime at a $\Delta F$ greater than that of the thin funnels. Indeed, as shown in figure \ref{fig:DC}, our numerics is consistent with the idea that the merger is a double-cone-like transition directly analogous to that described in \cite{Kol:2002xz,Kol:2005vy,Kalisch:2017bin,Kalisch:2018efd} for Kaluza-Klein black holes.  Thus even if further turning points are found, the associated phases are expected to have higher free energy.  Since the only remaining phase in figure \ref{fig:Full} is the droplet-plus-black-hole phase, one may thus expect that this is the relevant new phase new $\rho_{\mathrm{turn}}$.

\begin{figure}
\centerline{
\includegraphics[width=.5\textwidth]{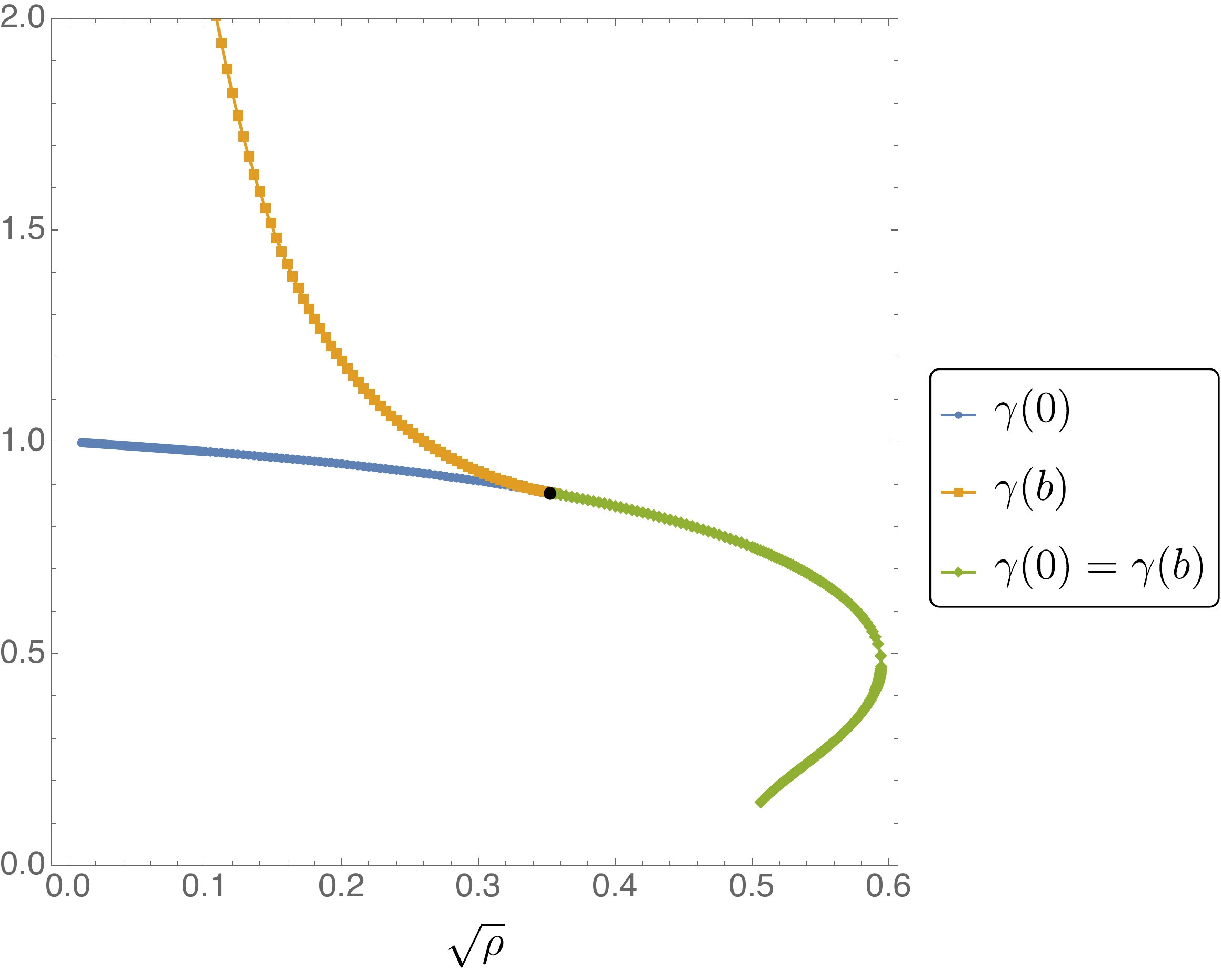}
}
\caption{The resdhift $\gamma(0)$ of the central geodesic and the maximum gravitational potential along the axis $\gamma(b)$ as a function of $\sqrt{\rho}$. These quantities are plotted respectively in blue and orange when they do not coincide.  But beyond a certain threshold (black dot) the maximum occurs at the origin and $\gamma(b)=\gamma(0)$.  In this regime (green) the central geodesic is unstable. Note that it includes the region near $\sqrt{\rho_{\mathrm{turn}}} \approx 0.5143$.}
\label{fig:DropletPotential}
\end{figure}

However, further study of the droplets near $\rho_{\mathrm{turn}}$ suggests that this is also unlikely. To understand this point, consider adding a small black hole to a droplet solution.  A sufficiently small black hole will act like a test particle and follow a geodesic, and by symmetry the static worldline at the origin is a geodesic in all droplet solutions.  But the lowest free energy must be dynamically stable, and the central geodesic is unstable near $\rho_{\mathrm{turn}}$.  This is shown by the results in figure \ref{fig:DropletPotential}, which plots the redshift $\gamma(0)=\sqrt{-g_{tt}(0)}$ at the origin as well as the height $\gamma(b)$ of any gravitational barrier that would stabilize the central geodesic with respect to perturbations along the axis of rotational symmetry.  Here we note that since $\sqrt{-g_{tt}}$ vanishes at horizons, it is bounded along this axis in a droplet phase.  We may thus define $b$ to be the point maximizing $\sqrt{-g_{tt}}$ and compute the associated barrier height $\gamma(b) = \sqrt{-g_{tt}}(b)$.  The issue is then that near $\sqrt{\rho_{\mathrm{turn}}} \approx 0.5143$ figure \ref{fig:DropletPotential} finds $\gamma(0)=\gamma(b)$ (i.e., that the gravitational potential along the axis is maximized at the origin),  so the central geodesic is unstable.

It is thus far from clear what new phase should dominate near $\sqrt{\rho_{\mathrm{turn}}}$.  This puzzle calls for further exploration of static phases or, even better, for dynamical real-time evolution of the system with time-dependent boundary conditions that increase $\rho$ across the threshold at $\rho_{\mathrm{turn}}$. Furthermore, while our numerics show the expected convergence (see appendix \ref{ap:1}) and also the expected behavior near the funnel/droplet merger (figure \ref{fig:DC}), it is difficult to exclude the suggestion that some subtle systematic effect might have led to small errors in our free energy computations that, when resolved will remove the need for this new phase.  It would thus also be useful for our results to be reproduced independently, perhaps using other computational frameworks.   Additional interesting directions to explore include the construction of droplet-plus-black-hole phases for both $d=4$ with SAdS boundaries and $d=3$ with BTZ boundaries, as in the latter case they would necessarily break the $SO(2,1)$ symmetry (see section \ref{BTZ}) used to analytically construct and analyze pure droplets and funnels in \cite{Hubeny:2009rc}.

\section*{Acknowledgements}
We thank Joan Camps, Harvey Reall, and Benson Way for related conversations.  D.M. would especially like to thank Veronika Hubeny and Mukund Rangamani for years of discussions of droplets and funnels.  Some of the ideas in this work no doubt date back to unpublished conversations related to the writing of \cite{Hubeny:2009rc}.  D.M.  was supported in part by the National Science Foundation under Grant Nos PHY1125915, PHY1504541, and PHY1801805, as well as by funds from the University of California. He also thanks DAMTP, Cambridge for their hospitality and Trinity College, Cambridge for support as a V.F.C. during critical stages of this work in 2013.  J.E.S. is supported in part by STFC grants PHY-1504541 and ST/P000681/1. This work used the DIRAC Shared Memory Processing system at the University of Cambridge, operated by the COSMOS Project at the Department of Applied Mathematics and Theoretical Physics on behalf of the STFC DiRAC HPC Facility (www.dirac.ac.uk). This equipment was funded by BIS National E- infrastructure capital grant ST/J005673/1, STFC capital grant ST/H008586/1, and STFC DiRAC Operations grant ST/K00333X/1. DiRAC is part of the National e-Infrastructure.
\begin{appendix}
\section{\label{ap:1}The differential operators $\mathcal{L}^{(i)}$, for $i\in\{0,1,2\}$:}
\begin{subequations}
\begin{multline}
\mathcal{L}^{(0)} = \left[2 (x-1)^8-(d+1) (1-x)^{d+5}+(d-1) (1-x)^{2 (d+1)}\right]\frac{\partial^2}{\partial x^2}+\\
\left\{[d (2 p-d+19)+2 (p-6)] (1-x)^{d+4}-(d-1) (1-x) (d+2 p+6) (1-x)^{2 d}\right.\\
\left.-2 (1-x)^7 (d+2 p+6)\right\}\frac{\partial}{\partial x}+\{d^2 (1-x)^d \left[(p+3) (1-x)^d-(p+1) (x-1)^3\right]+\\
(p+2) \left[2 (p+3) (x-1)^6-(p+3) (1-x)^{2 d}-(p-15) (1-x)^{d+3}\right]+\\
d \left[(p+1) (p+3) (1-x)^{2 d}-[p(p+18)+33] (1-x)^{d+3}+2 (p+3) (x-1)^6\right]\}\,,
\end{multline}
\begin{multline}
\mathcal{L}^{(1)}=(1-x)^2 \{2 (1-x)^4-(1-x) [d (x^2-2 x+2)+(x-2) x] (1-x)^d\\
+2 (d-1) (1-x)^{2 d}\}\frac{\partial^2}{\partial x^2}-(1-x) \left\{(1-x)^{d+1} [d^2 (x^2-2x+2)-2 p d(x^2-2 x+2)\right. \\
\left.-19 d (x-2) x-28 d-2 (p-6) (x-2) x+20]+2 (d-1) (d+2 p+6) (1-x)^{2 d}\right. \\
\left.+2 (x-1)^4 (d+2 p+2)\right\}\frac{\partial}{\partial x}+(1-x)^d \{d^2 [2 (p+3) (1-x)^d+(1-x) (p x^2-2 p x+3p+x^2-2 x+5)]\\
-d (1-x) \left[p^2 (1-x)^2+2 p (9 x^2-18 x+13)-33 (2-x) x+49\right]+2 d (p+1) (p+3)(1-x)^d \\
-(p+2) \left[2 (p+3) (1-x)^d+(1-x) \left(p x^2-2 p x+p-15 x^2+30 x-21\right)\right]\}\,,
\end{multline}
and
\begin{multline}
\mathcal{L}^{(2)} = (d-1) (1-x)^{d+2} \left[(1-x)^d-(1-x)\right]\frac{\partial^2}{\partial x^2}-(d-1) (1-x)^{d+1} \left[(d+2 p+6) (1-x)^d\right.\\
\left.+(1-x) (d-2 p-8)\right]\frac{\partial}{\partial x}+(d-1) (1-x)^d \left[(p+3) (d+p+2) (1-x)^d+2 (d-3) (p+2) (1-x)\right]\,.
\end{multline}
\end{subequations}
\end{appendix}
\section{\label{ap:2} Convergence tests}
Of all the quantities we computed in the main text, the energy and entropy variations with respect to the uniform funnel are the most difficult ones to extract. Note that once we know these two quantities, we can easily compute the corresponding free energy variation via standard thermodynamic relations. The entropy variation turns out to be very accurate to determine, and its numerical error is always below the $10^{-8}\%$. So we focus on the energy extraction. We will also study convergence keeping the number of points in the $x$ and $y$ directions equal. That is to say, we set $n_x=n_y=n$. For each value of $n$ we determine an energy variation $\Delta E_n$. To study its convergence we monitor the quantity
\begin{equation}
\delta E_n = \left|1-\frac{\Delta E_n}{\Delta E_{n+10}}\right|\,.
\end{equation}

\begin{figure}
\centerline{
\includegraphics[width=\textwidth]{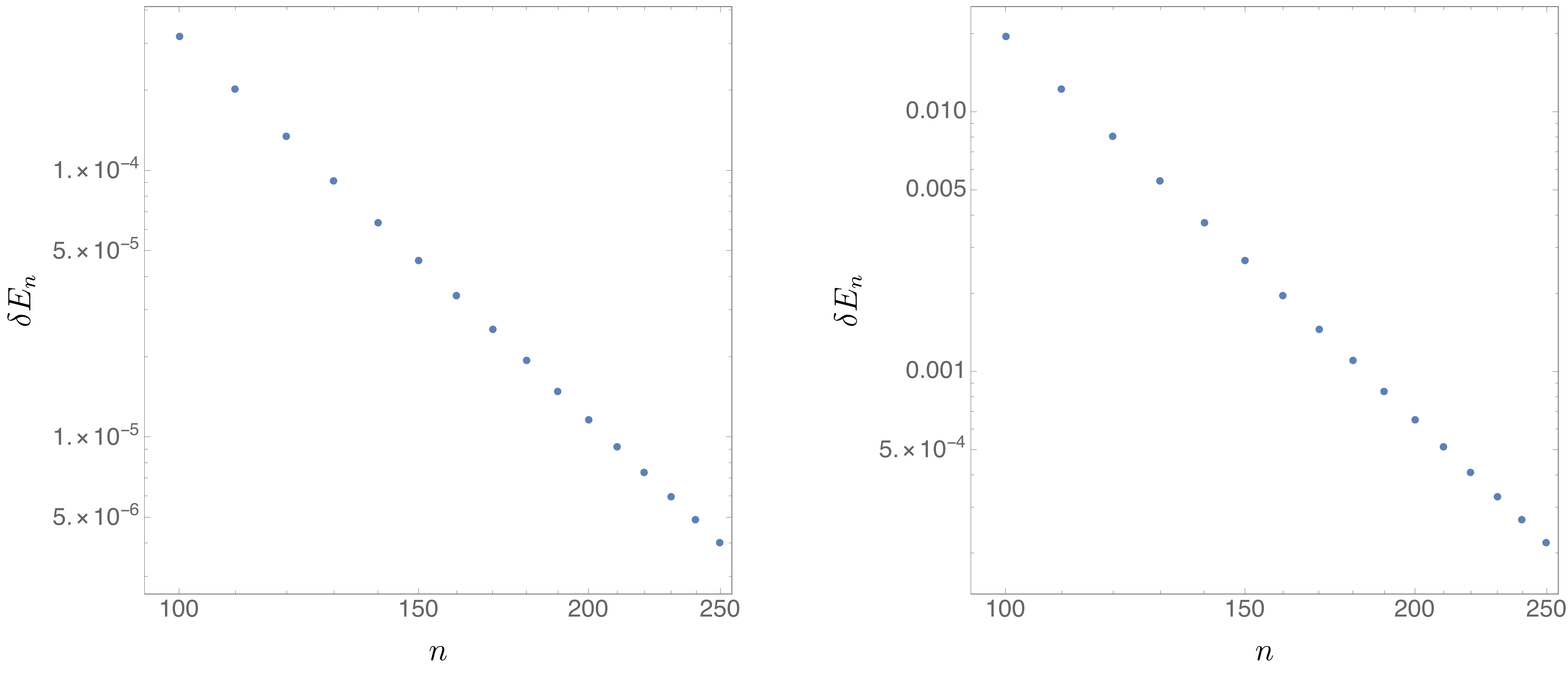}
}
\caption{$\delta E_n$ as a function of $n$ for the droplet (left panel) and funnel (right panel) phases. In both cases we used $\sqrt{\rho}=0.312$.}
\label{fig:conv}
\end{figure}

Our results indicate a power law convergence for the energy, which is not surprising given the irrational decays we found in Eqs.~(\ref{eqs:irrfunnels}) and Eqs.~(\ref{eqs:expansiony1}). For reference, we plot in Fig.~\ref{fig:conv} a convergence plot for the droplet (left panel) and funnel (right panel) for $\sqrt{\rho}=0.312$. For the droplets we find $\Delta E_n \propto n^{-4.7(8)}$ while for the funnels $\Delta E_n \propto n^{-4.9(0)}$. Other values of $\rho$ exhibit similar convergence properties.

In all plots in our manuscript, we used $n=250$. We have also numerically checked the first law of thermodynamics shown in Eq.~(\ref{eq:firstlaw}), and find that it is satisfied to $10^{-3}\%$ level.

\bibliographystyle{JHEP}
\bibliography{FDBib}

\end{document}